\documentclass[subeqn]{article}

  \usepackage[T1]{fontenc}
  \usepackage{amsmath,amssymb,mathrsfs}
  \usepackage[left=4cm, right=4cm, top=3cm, bottom=3cm]{geometry}
  \usepackage{enumitem}
\usepackage[toc,page]{appendix}

\usepackage[ansinew]{inputenc}
\usepackage{cases}
\usepackage{cases}
\usepackage{array}
\usepackage{mathenv}

\usepackage{graphicx,color}
\usepackage{eurosym}
\usepackage{amsmath,amsfonts,amssymb,amsthm}
\usepackage{mathrsfs}
\usepackage{wasysym}
\usepackage{oldstyle}
\usepackage[T1]{fontenc}
\usepackage{pb-diagram}

  \theoremstyle{plain}
   \newtheorem{theorem}{Theorem}
   \newtheorem{lemma}{Lemma}
   \newtheorem{proposition}{Proposition}
  \theoremstyle{definition}
  \newtheorem{definition}{Definition}
  \theoremstyle{remark}
    \newtheorem{remark}{Remark}
   \newtheorem{notation}{Notation}

\usepackage{hyperref}
\title{\textbf{The positronium in a mean-field approximation of quantum electrodynamics.}}
\author{Sok J\'er\'emy\\
 Ceremade, UMR 7534, Universit\'e Paris-Dauphine,\\
  Place du Mar\'echal de Lattre de Tassigny,\\
  75775 Paris Cedex 16, France.\\ \\
}%

\usepackage{xspace}
\newcommand{\ket}[1]{\ensuremath{|#1\rangle}\xspace}
\newcommand{\bra}[1]{\ensuremath{\langle #1|}\xspace}
\newcommand{\psh}[2]{\ensuremath{\langle #1\,,\,#2\rangle}\xspace}

\newcommand{\ssum}{\ensuremath{\displaystyle\sum}}
\newcommand{\dint}{\ensuremath{\displaystyle\int}}
\newcommand{\diint}{\ensuremath{\displaystyle\iint}}

\newcommand{\g}{\ensuremath{\gamma}}
\newcommand{\G}{\ensuremath{\Gamma}}

\newcommand{\pvac}{\ensuremath{\overline{\boldsymbol{\pi}}}}
\newcommand{\pvt}{\ensuremath{\boldsymbol{\pi}}}

\newcommand{\ns}[2]{\ensuremath{\lVert#2\rVert_{\mathfrak{S}_{#1}}}}
\newcommand{\nlp}[2]{\ensuremath{\lVert#2\rVert_{L^{#1}}}}

\newcommand{\nqq}[1]{\ensuremath{\lVert#1\rVert_{\text{Ex}}}}
\newcommand{\nb}[1]{\ensuremath{\lVert#1\rVert_{\mathcal{B}}}}

\newcommand{\ov}[1]{\ensuremath{\overline{#1}}}
\newcommand{\un}[1]{\ensuremath{\underline{#1}}}

\newcommand{\llo}{\ensuremath{\log(\Lambda)}}

\newcommand{\ph}{\ensuremath{\varphi}}

\newcommand{\wh}[1]{\ensuremath{\widehat{#1}}}
\newcommand{\la}{\ensuremath{\lambda}}
\newcommand{\La}{\ensuremath{\Lambda}}
\newcommand{\ttr}{\ensuremath{\mathrm{Tr}}}

\newcommand{\wt}[1]{\ensuremath{\widetilde{#1}}}
\newcommand{\D}{\ensuremath{\mathcal{D}^0}}
\newcommand{\dd}{\ensuremath{\mathrm{d}}}
\newcommand{\hl}{\ensuremath{\mathfrak{H}_\Lambda}}

\newcommand{\PP}{\ensuremath{\mathcal{P}^0_-}}
\newcommand{\PPP}{\ensuremath{\mathcal{P}^0_+}}

\newcommand{\Cha}{\ensuremath{\mathrm{C}}}

\newcommand{\eps}{\ensuremath{\varepsilon}}

\newcommand{\om}{\ensuremath{\omega}}
\newcommand{\ed}[1]{\ensuremath{\widetilde{E}\left(#1\right)}}

\newcommand{\CC}{\ensuremath{\mathbb{C}^4}}
\newcommand{\RR}{\ensuremath{\mathbb{R}^3}}

\begin{document}
\maketitle

\abstract{The Bogoliubov-Dirac-Fock (BDF) model is a no-photon, mean-field approximation of quantum electrodynamics. It describes relativistic electrons in the Dirac sea.
In this model, a state is fully characterized by its one-body density matrix, an infinite rank nonnegative operator. We prove the existence of the positronium, the bound state of an electron and  a positron, represented by a critical point of the energy functional in the absence of external field. This state is interpreted as the ortho-positronium, where the two particles have parallel spins.}

\tableofcontents

\newpage
\section{Introduction and main results}

\begin{center}\textsc{The Dirac operator}
\end{center}

In relativistic quantum mechanics, the kinetic energy of an electron is described by the so-called \emph{Dirac operator} $D_0$. Its expression is \cite{Th}:
\begin{equation}\label{dirac_op}
 D_0:=m_ec^2\beta-i\hbar c\ssum_{j=1}^3\alpha_j \partial_{x_j}
\end{equation}
where $m_e$ is the (bare) mass of the electron, $c$ the speed of light and $\hbar$ the reduced Planck constant, $\beta$ and the $\alpha_j$'s are $4\times 4$ matrices defined as follows:
\begin{equation*}\label{beta_alpha}
 \beta:=\begin{pmatrix} 
        \mathrm{Id}_{\mathbb{C}^2} & 0\\ 0 & -\mathrm{Id}_{\mathbb{C}^2}
        \end{pmatrix},\ \alpha_j \begin{pmatrix} 
         0 & \sigma_j \\ \sigma_j & 0
        \end{pmatrix},\ j\in\{1,2,3\}
\end{equation*}
\begin{equation*}
 \sigma_1:=\begin{pmatrix} 
         0 & 1\\ 1 & 0
        \end{pmatrix},\ \sigma_2:=\begin{pmatrix} 
         0 & -i\\ i & 0
        \end{pmatrix},\ \sigma_3\begin{pmatrix} 
         1 & 0 \\ -1 & 0
        \end{pmatrix}.
\end{equation*}
 The Dirac operator acts on spinors \emph{i.e.} square-integrable $\mathbb{C}^4$-valued functions:
\begin{equation}\label{space_one_electron}
 \mathfrak{H}:=L^2\big(\RR,\CC\big).
\end{equation}
It corresponds to the Hilbert space associated to one electron. The operator $D_0$ is self-adjoint on $\mathfrak{H}$ with domain $H^1(\RR,\CC)$, but contrary to $-\Delta/2$ in quantum mechanics, it is unbounded from below. 

Indeed its spectrum is $\sigma(D_0)=(-\infty,m_ec^2]\cup[m_e c^2,+\infty)$. Dirac postulated that all the negative energy states are already occupied by "virtual electrons", with one electron in each state, and that the uniform filling is unobservable to us. 
Then, by Pauli's principle real electrons can only have a positive energy. 

It follows that the relativistic vacuum, composed by those negatively charged virtual electrons, is a polarizable medium that reacts to the presence of an external field. This phenomenon is called the \emph{vacuum polarization}. 

If one turns on an external field that gets strong enough, it leads to a transition of an electron of the Dirac sea from a negative energy state to a positive one. The resulting system -- an electron with positive energy plus a hole in the Dirac sea -- is interpreted as an electron-positron pair. Indeed the absence of an electron in the Dirac sea is equivalent to the addition of a particle with same mass and opposite charge: the positron.

Its existence was predicted by Dirac in 1931. Although firstly observed in 1929 independently by Skobeltsyn and Chung-Yao Chao, it was recognized in an experiment lead by Anderson in 1932.

\medskip

\begin{center}
\textsc{Charge conjugation} 
\end{center}

Following Dirac's ideas, the free vacuum is described by the negative part of the spectrum $\sigma(D_0)$:
\[
P^0_-=\chi_{(-\infty,0)}(D_0).
\]
The correspondence between negative energy states and positron states is given by the \emph{charge conjugation} $\Cha$ \cite{Th}. This is an antiunitary operator that maps $\text{Ran}\,P^0_{-}$ onto $\text{Ran}(1-P^0_{-})$. In our convention \cite{Th} it is defined by the formula:
\begin{equation}\label{chargeconj}
\forall\,\psi\in L^2(\RR),\ \Cha\psi(x)=i\beta\alpha_2\overline{\psi}(x),
\end{equation}
where $\overline{\psi}$ denotes the usual complex conjugation. More precisely:
\begin{equation}\label{chargeconjprec}
\Cha\cdot \begin{pmatrix}\psi_1\\ \psi_2\\ \psi_2\\\psi_4\end{pmatrix}=\begin{pmatrix}\overline{\psi}_4\\ -\overline{\psi}_3\\ -\overline{\psi}_2\\\overline{\psi}_1\end{pmatrix}.
\end{equation}
In our convention it is also an \emph{involution}: $\Cha^2=\text{id}$. An important property is the following:
\begin{equation}\label{denspsi}
\forall\,\psi\in\,L^2,\forall\,x\in\mathbb{R}^3,\ |\Cha \psi(x)|^2=|\psi(x)|^2.
\end{equation}

\medskip
\begin{center}
\textsc{Positronium} 
\end{center}

The positronium is the bound state of an electron and a positron. This system was independently predicted by Anderson and Mohorovi$\check{\mathrm{c}}$i\'c in 1932 and 1934 and was experimentally observed for the first time in 1951 by Martin Deutsch. 

It is unstable: depending on the relative spin states of the positron and the electron, its average lifetime in vacuum is 125 ps (para-positronium) or 142 ns (ortho-positronium) (see \cite{karsh}). 

In this paper, we are looking for a positronium state within the Bogoliubov-Dirac-Fock (BDF) model: the state we found can be interpreted as the ortho-positronium where the electron and positron have parallel spins. Our main results are Theorem \ref{critic} and \ref{non_rel}. In our state, the wave function of the real electron and that of the virtual electron defining the positronium are charge conjugate of each other.

\medskip
\begin{center}
\textsc{BDF model} 
\end{center}

The BDF model is a no-photon approximation of quantum electrodynamics (QED) which was introduced by Chaix and Iracane in 1989 \cite{CI}, and studied in many papers \cite{stab,ptf,Sc,mf,at,gs,sok}.

It allows to take into account real electrons together with the Dirac vacuum in the presence of an external field. 

This is a Hartree-Fock type approximation in which a state of the system "vacuum $+$ real electrons" is given by an infinite Slater determinant $\psi_1\wedge\psi_2\wedge \cdots$. Equivalently, such a state is represented by the projector onto the space spanned by the $\psi_j$'s: its so-called one-body density matrix. For instance $P^0_-$ represents the free Dirac vacuum. 

Here we just give main ideas of the derivation of the BDF model from QED, we refer the reader to \cite{CI,ptf,mf} for full details.

\begin{remark}
To simplify the notations, we choose relativistic units in which, the mass of the electron $m_e$, the speed of light $c$ and $\hbar$ are set to $1$.
\end{remark}
Let us say that there is an external density $\nu$, \emph{e.g.} that of some nucleus and let us write $\alpha>0$ the so-called \emph{fine structure constant} (physically $e^2/(4\pi\eps_0\hbar c)$, where $e$ is the elementary charge and $\eps_0$ the permittivity of free space). 

The starting point is the (complicated) Hamiltonian of QED $\mathbb{H}_{\text{QED}}$ that acts on the Fock space of the electron $\mathcal{F}_{\text{elec}}$ \cite{Th}. The (formal) difference between the infinite energy of a Hartree-Fock state $\Omega_P$ and that of $\Omega_{P^0_-}$, state of the free vacuum taken as a reference state, gives a function of the reduced one-body density matrix $Q:=P-P^0_-$.


It can be shown that a projector $P$ is the one-body density matrix of a Hartree-Fock state in $\mathcal{F}_{\text{elec}}$ \emph{iff} $P-P^0_-$ is Hilbert-Schmidt, that is compact such that its singular values form a sequence in $\ell^2$. 

To get a well-defined energy, one has to impose an ultraviolet cut-off $\Lambda>0$: we replace $\mathfrak{H}$ by its subspace
\[
 \hl:=\big\{ f\in\mathfrak{H},\ \text{supp}\,\wh{f}\subset B(0,\La)\big\}.
\]
This procedure gives the BDF energy introduced in \cite{CI} and studied for instance in \cite{ptf,Sc}.
\begin{notation}
Our convention for the Fourier transform $\mathscr{F}$ is the following
\[
\forall\,f\in L^1(\RR),\ \wh{f}(p):=\frac{1}{(2\pi)^{3/2}}\dint f(x)e^{-ixp}dx.
\]
\end{notation}
Let us notice that $\hl$ is invariant under $D_0$ and so under $P^0_-$.

For the sake of clarity, we will emphasize the ultraviolet cut-off and write $\Pi_\La$ for the orthogonal projection onto $\hl$: $\Pi_\La$ is the following Fourier multiplier
\begin{equation}
 \Pi_\La:=\mathscr{F}^{-1}\chi_{B(0,\Lambda)}\mathscr{F}.
\end{equation}

By means of a thermodynamical limit, Hainzl \emph{et al.} showed in \cite{mf} that the formal minimizer and hence the reference state should not be given by $\Pi_\La P^0_-$ but by another projector $\PP$ in $\hl$ that satisfies the self-consistent equation \underline{\emph{in $\hl$}}:
\begin{equation}\label{PP_self}
 \left\{ \begin{array}{ccl}
          \PP-\tfrac{1}{2}&=&-\text{sign}\big(\D\big),\\
          \D&=&D_0 -\dfrac{\alpha}{2}\dfrac{(\PP-\tfrac{1}{2})(x-y)}{|x-y|}
         \end{array}
\right.
\end{equation}
We have $\PP=\chi_{(-\infty,0)}(\D)$. 

In $\mathfrak{H}$, the operator $\D$ coincides with a bounded, matrix-valued Fourier multiplier whose kernel is $\hl^{\perp}\subset \mathfrak{H}$.

\medskip

The resulting BDF energy $\mathcal{E}^\nu_{\text{BDF}}$ is defined on Hartree-Fock states represented by their one-body density matrix $P$:
\[
\mathscr{N}:=\big\{P\in\mathcal{B}(\hl),\ P^*=P^2=P,\ P-\PP\in\mathfrak{S}_2(\hl)\big\}.
\]

This energy depends on three parameters: the fine structure constant $\alpha>0$, the cut-off $\La>0$ and the external density $\nu$. We assume that $\nu$ has finite \emph{Coulomb energy}, that is 
\begin{equation}
D(\nu,\nu):=4\pi\underset{\RR}{\dint}\frac{|\wh{\nu}(k)|^2}{|k|^2}dk.
\end{equation}

\begin{remark}
The Coulomb energy coincides with $\underset{\RR\times\RR}{\iint}\frac{\nu(x)^*\nu(y)}{|x-y|}dxdy$ whenever this integral is well-defined.
\end{remark}

\begin{remark}
The operator $\D$ was previously introduced by Lieb \emph{et al.} in \cite{ls} in another context in the case $\alpha\llo$ small. 
\end{remark}

\begin{notation}
We recall that $\mathcal{B}(\hl)$ is the set of bounded operators and $\mathfrak{S}_p(\hl)$ the set of compact operators whose singular values form a sequence in $\ell^p$ \cite[Appendix IX.4 Vol II]{ReedSim}, \cite{Sim} ($p\ge1$). In particular $\mathfrak{S}_{\infty}(\hl)$ is the set $\text{Comp}(\hl)$ of compact operators.
\end{notation}

\begin{notation}
Throughout this paper we write 
\begin{equation}
m=\inf \sigma \big(|\D|\big)\ge 1,
\end{equation}
and
\begin{equation}
\PPP:=\Pi_\La-\PP=\chi_{(0,+\infty)}(\D).
\end{equation}
The same symmetry holds for $\PP$ and $\PPP$: the charge conjugation $\Cha$ maps $\text{Ran}\,\PP$ onto $\text{Ran}\,\PPP$.
\end{notation}

\medskip
\begin{center}
\textsc{Minimizers and critical points}
\end{center}

The charge of a state $P\in\mathscr{N}$ is given by the so-called $\PP$-trace of $P-\PP$
\[
\ttr_{\PP}\big(P-\PP\big):=\ttr\big(\PP (P-\PP)\PP+\PPP(P-\PP)\PPP\big).
\]
This trace is well defined as we can check from the formula \cite{ptf}
\begin{equation}\label{eqq}
(P-\PP)^2=\PPP (P-\PP)\PPP-\PP(P-\PP)\PP.
\end{equation}
Minimizers of the BDF energy with charge constraint $N\in\mathbb{N}$ corresponds to ground states of a system of $N$ electrons in the presence of an external density $\nu$. 

The problem of their existence was studied in several papers \cite{at,sok,sokd}. In \cite{at}, Hainzl \emph{et al.} proved that it was sufficient to check binding inequalities and showed existence of ground states in the presence of an external density $\nu$, provided that $N-1<\int \nu$, under technical assumptions on $\alpha,\La$. 

In \cite{sok}, we proved that, due to the vacuum polarization, there exists a minimizer for $\mathcal{E}^0_{\text{BDF}}$ with charge constraint $1$: in other words an electron can bind alone in the vacuum without any external charge (still under technical assumptions on $\alpha,\La$).

In \cite{sokd}, the effect of charge screening is studied: due to vacuum polarization, the observed charge of a minimizer $P\neq \PP$ is different from its real charge $\ttr_{\PP}(P-\PP)$.

\medskip

Here we are looking for a positronium state, that is an electron and a positron in the vacuum without any external density. So we have to study $\mathcal{E}^0_{\text{BDF}}$ on 
\begin{equation}
\mathscr{M}:=\Big\{P\in\mathscr{N},\ \ttr_{\PP}\big(P-\PP\big)=0\Big\}.
\end{equation}
From a geometrical point of view $\mathscr{M}$ is a Hilbert manifold and $\mathcal{E}^0_{\text{BDF}}$ is a differentiable map on $\mathscr{M}$ (Propositions \ref{manim} and \ref{gragra}).

We thus seek a critical point on $\mathscr{M}$, that is some $P\in\mathscr{M},\ P\neq\PP$ such that $\nabla \mathcal{E}^0_{\text{BDF}}(P)=0$.
We also must ensure that this is a positronium state. A good candidate is a projector $P$ that is obtained from $\PP$ by substracting a state $\psi_-\in\,\text{Ran}\,\PP$ and adding a state $\psi_+\in\,\text{Ran}\,\PPP$, that is
\begin{equation}\label{imagine}
 P=\PP+\ket{\psi_+}\bra{\psi_+}-\ket{\psi_-}\bra{\psi_-}.
\end{equation}
But there is no reason why such a projector would be a critical point. If it were that would mean that there exists a positronium state in which, apart from the excitation of the virtual electron giving the electron-positron pair, the vacuum is not polarized.

Keeping \eqref{imagine} in mind, we identify a subset $\mathscr{M}_{\mathscr{C}}\subset \mathscr{M}$, made of $\Cha$-symmetric states.
\begin{definition}
The set $\mathscr{M}_{\mathscr{C}}$ of $\Cha$-symmetric states is defined as:
\begin{equation}\label{def_mc}
\mathscr{M}_{\mathscr{C}}=\{ P\in\mathscr{M},\ -\Cha (P-\PP)\Cha=P-\PP\}.
\end{equation}
\end{definition}

\begin{remark}
 Let $P\in\mathscr{M}_{\mathscr{C}}$. As $-\Cha (\PP-\PPP)\Cha=\PP-\PPP$, writing
 \[
 P-\PP=\tfrac{1}{2}\big(P-(\Pi_\La-P)-\PP+\PPP\big),
 \] there holds:
 \begin{equation}\label{chaisom}
  P\in\mathscr{M}_{\mathscr{C}}\Rightarrow P+\Cha P\Cha=\Pi_\La,
 \end{equation}
that is 
\begin{equation*}
\forall\,P\in\mathscr{M}_{\mathscr{C}},\ \Cha:\text{Ran}\,P\to \text{Ran}(\Pi_\La-P)\text{\ is\ an\ isometry.}
 \end{equation*}
\end{remark}

\medskip

The set $\mathscr{M}_{\mathscr{C}}$ has fine properties: this is a submanifold, invariant under the gradient flow of $\mathcal{E}^0_{\text{BDF}}$ (Proposition \ref{manicsym}). Moreover it has two connected components $\mathscr{E}_1$ and $\mathscr{E}_{-1}$ (Proposition \ref{conn}). In particular, any extremum of the BDF energy restricted to $\mathscr{M}_{\mathscr{C}}$ is a critical point on $\mathscr{M}$.

So we are lead to seek a minimizer over each of these connected components: the first ($\mathscr{E}_1$) gives $\PP$, which is the global minimizer over $\mathscr{N}$, but the second gives a non-trivial critical point. It corresponds to the positronium and is a perturbation of a state which can be written as in \eqref{imagine}.

Our main Theorems are the following:
\begin{theorem}\label{critic}
There exist $\alpha_0,\La_0,L_0>0$ such that if $\alpha\le \alpha_0,\La^{-1}\le \La_0^{-1},$ and $\alpha\log(\La)\le L_0$, then there exists a minimizer of $\mathcal{E}_{\mathrm{BDF}}^0$ over $\mathscr{E}_{-1}$. Moreover we have
\[
E_{1,1}:=\inf\{ \mathcal{E}_{\mathrm{BDF}}^0(P),\ P\in\mathscr{E}_{-1}\}\le 2m+\frac{\alpha^2 m}{g'_1(0)^2}E_{\mathrm{CP}}+ \mathcal{O}(\alpha^3),
\]
where $E_{\mathrm{CP}}<0$ is the Choquard-Pekar energy defined as follows \cite{L}:
\begin{equation}
 E_{\mathrm{CP}}=\inf\Big\{\nlp{2}{\nabla \phi}^2-D\big(|\phi|^2,|\phi|^2\big),\ \phi\in L^2(\mathbb{R}^3),\ \nlp{2}{\phi}=1\Big\}.
\end{equation}
\end{theorem}

\begin{theorem}\label{foform}
Under the same assumptions as in Theorem \ref{critic}, let $\ov{P}$ be a minimizer for $E_{1,1}$. Then there exists an anti-unitary map $A\in\mathbf{A}(\hl)$, and $\text{P}^0_{1,1}$ of form \eqref{imagine} such that 
\begin{equation}\label{form_minim}
\begin{array}{|l}
\ov{P}=e^A \text{P}_{1,1}^0 e^{A},\\
e^A\psi_\eps=\psi_\eps,\ \eps\in\{+,-\}\text{\ and\ }
\psi_-=\Cha \psi_+,\\
A=\big[\big[A,\PP\big],\PP\big]\in\mathfrak{S}_2(\hl),\ 
\ns{2}{A}\apprle \alpha,\\
\text{and\ }\Cha A \Cha =A.
\end{array}
\end{equation}
Moreover, the following holds:
\begin{equation}\label{lowup}
E_{1,1}= 2m+\frac{\alpha^2 m}{g'_1(0)^2}E_{\mathrm{CP}}+ \mathcal{O}(\alpha^3).
\end{equation}
\end{theorem}

We emphasize that $\psi_+$ \emph{does not} represent the electron state.

\begin{theorem}\label{non_rel}
Under the same assumptions as in Theorem \ref{critic}, let $\ov{P}$ be a minimizer for $E_{1,1}$ and $Q_0=\ov{P}-\PP$. Let $\pvac$ be
\begin{equation}
\pvac:=\chi_{(-\infty,0)}\big(\Pi_\La D_{Q_0}\Pi_\La\big).
\end{equation}
Then there holds $\mathrm{Ran}\,(\Pi_\La -\pvac)\cap \mathrm{Ran}\,\ov{P}=\mathbb{C}\psi_e$. The unitary wave function $\psi_e$ satisfies the equation
\begin{equation}
D_{Q_0}\psi_e=\mu_e \psi_e,
\end{equation}
where $\mu_e$ is some constant
\[
K_0\alpha^2\le m-\mu_e\le K_1\alpha^2,\ K_0,K_1>0.
\]
By $\Cha$-symmetry $\psi_v:=\Cha \psi_e$ satisfies $D_{Q_0}\psi_v=-\mu_e \psi_v$, and we have
\begin{equation}
\ov{P}=\pvac+\ket{\psi_e}\bra{\psi_e}-\ket{\psi_v}\bra{\psi_v}.
\end{equation}
Moreover the following holds. We split $\psi_e$ into upper spinor $\ph_e\in L^2(\RR,\mathbb{C}^2)$ and lower spinor $\chi_e \in L^2(\RR,\mathbb{C}^2)$ and scale $\ph_e$ by $\la:=\tfrac{g'_1(0)^2}{\alpha m}$:
\[
\wt{\ph}_e(x):=\la^{3/2}\ph_e\big(\la x\big).
\]
Then in the non-relativistic limit $\alpha\to 0$ (with $\alpha\llo$ kept small), the lower spinor $\chi_e$ tends to $0$ and, up to translation, $\wt{\ph}_e$ tends to a Pekar minimizer.
\end{theorem}

\begin{remark}
As $\psi_e$ and $\psi_v=\Cha \psi_e$ have antiparallel spins, the state $\ov{P}$ represents one electron in state $\psi_e$ and the absence of one electron in state $\psi_v$ in the Dirac sea, that is an electron and a positron with \emph{parallel spins}.
\end{remark}

\begin{remark}
 To prove that $\wt{\ph}_e$ tends to a Pekar minimizer up to translation, it suffices to prove that its Pekar energy tends to $E_{\text{CP}}$ \cite{L}.
\end{remark}

\begin{notation}
Throughout this paper we write $K$ to mean a constant independent of $\alpha,\La$. Its value may differ from one line to the other. We also use the symbol $\apprle$: $0\le a\apprle b$ means there exists $K>0$ such that $a\le Kb$.
\end{notation}

\medskip

\begin{center}
\textsc{Remarks and notations about $\D$}
\end{center}

$\D$ has the following form \cite{mf}:
\begin{equation}\label{D_form}
 \D=g_0(-i\nabla)\beta -i\boldsymbol{\alpha}\cdot \frac{\nabla}{|\nabla|}g_1(-i\nabla)
\end{equation}
where $g_0$ and $g_1$ are smooth radial functions on $B(0,\La)$ and $\boldsymbol{\alpha}=(\alpha_j)_{j=1}^3$. Moreover we have:
\begin{equation}
 \forall\,p\in B(0,\La),\ 1\le g_0(p),\text{\ and\ }|p|\le g_1(p)\le |p|g_0(p).
\end{equation}
\begin{notation}
For $\alpha\llo$ sufficiently small, we have $m=g_0(0)$ \cite{LL,sok}.
\end{notation}

\begin{remark}
In general the smallness of $\alpha$ is needed to ensure technical estimates hold. The smallness of $\alpha\llo$ is needed to get estimates of $\D$: in this case $\D$ can be obtained by a fixed point scheme \cite{mf,LL}, and we have \cite[Appendix A]{sok}:
\begin{equation}\label{estim_g}
\begin{array}{c}
g'_0(0)=0,\ \text{and}\ \nlp{\infty}{g'_0},\nlp{\infty}{g_0''}\le K\alpha\\
\nlp{\infty}{g'_1-1}\le K\alpha\llo\le \tfrac{1}{2}\ \text{and}\ \nlp{\infty}{g_1''}\apprle 1.
\end{array}
\end{equation}
\end{remark}


\ \newline

\noindent\textit{Acknowledgment} The author wishes to thank \'Eric S\'er\'e for useful discussions and helpful comments. This work was partially supported by the Grant ANR-10-BLAN 0101 of the French Ministry of Research.

\section{Description of the model}

\subsection{The BDF energy}


\begin{definition}
Let $\alpha>0,\La>0$ and $\nu\in\mathcal{S}'(\RR)$ a generalized function with $D(\nu,\nu)<+\infty$. The BDF energy $\mathcal{E}^0_{\text{BDF}}$ is defined on $\mathscr{N}$ as follows: for $P\in\mathscr{N}$ we write $Q:=P-\PP$ and
\begin{equation}\label{formule_bdf}
\left\{\begin{array}{l}
\mathcal{E}^0_{\text{BDF}}(Q)=\ttr_{\PP}\big(\D Q \big)-\alpha D(\rho_Q,\nu)+\dfrac{\alpha}{2}\Big(D(\rho_Q,\rho_Q)-\nqq{Q}^2\Big),\\
\forall\,x,y\in\RR,\ \rho_Q(x):=\ttr_{\mathbb{C}^4}\big(Q(x,x)\big),\ \nqq{Q}^2:=\diint\frac{|Q(x,y)|^2}{|x-y|}dxdy,
\end{array}\right.
\end{equation}
where $Q(x,y)$ is the integral kernel of $Q$. 
\end{definition}
\begin{remark}
The term $\ttr_{\PP}\big(\D Q \big)$ is the kinetic energy, $-\alpha D(\rho_Q,\nu)$ is the interaction energy with $\nu$. The term $\dfrac{\alpha}{2}D(\rho_Q,\rho_Q)$ is the so-called \emph{diract term} and $-\dfrac{\alpha}{2}\nqq{Q}^2$ is the \emph{exchange term}.
\end{remark}

\noindent 1. Let us see that this function is well-defined and more generally that formula \eqref{formule_bdf} is well-defined whenever $Q$ is $\PP$-trace-class \cite{ptf,at}.

\medskip

\noindent -- We start by defining this notion. For any $\eps,\eps'\in\{+,-\}$ and $A\in\mathcal{B}(\hl)$, we write
\[
A^{\eps,\eps'}:=\mathcal{P}^0_{\eps}A\mathcal{P}^0_{\eps'}.
\]
The set $\mathfrak{S}_1^{\PP}$ of $\PP$-trace class operator is the following Banach space:
\begin{equation}
\mathfrak{S}_1^{\PP}=\big\{Q\in\mathfrak{S}_2(\hl),\ Q^{++},Q^{--}\in\mathfrak{S}_1(\hl)\big\},
\end{equation} 
with the norm
\begin{equation}
\lVert Q\rVert_{\mathfrak{S}_1^{\PP}}:=\ns{2}{Q^{+-}}+\ns{2}{Q^{-+}}+\ns{1}{Q^{++}}+\ns{1}{Q^{--}}.
\end{equation}


We have $\mathscr{N}\subset \PP+\mathfrak{S}_1^{\PP}$ thanks to Eq. \eqref{eqq}. The closed convex hull of $\mathscr{N}-\PP$ in the $\mathfrak{S}_1^{\PP}$-topology gives
\[
\mathcal{K}:=\big\{Q\in\mathfrak{S}_1^{\PP}(\hl),\ Q^*=Q,\ -\PP\le Q\le \PPP\big\}
\]
and we have \cite{ptf,Sc}: $\forall\,Q\in \mathcal{K},\ Q^2\le Q^{++}-Q^{--}.$

\noindent -- For $Q$ in $\mathfrak{S}_1^{\PP}$, we show $\mathcal{E}^\nu_{\text{BDF}}(Q)$ is well defined. 
We have
\[
\PP(\D Q)\PP=-|\D|Q^{--}\in\,\mathfrak{S}_1(\hl),\ \text{because}\, |\D|\in\mathcal{B}(\hl),
\]
this proves that the kinetic energy is defined.

\noindent -- Thanks to the Kato-Seiler-Simon inequality \cite[Chapter 4]{Sim}, the operator $Q$ is locally trace-class:
\[
 \forall\,\phi\in \mathbf{C}^\infty_0(\RR),\ \phi \Pi_\La\in\mathfrak{S}_2\text{\ so\ }\phi Q \phi=\phi\Pi_\La Q\phi\in\mathfrak{S}_1(L^2(\RR)).
\]
We recall this inequality states that for all $2\le p\le \infty$ and $d\in\mathbb{N}$, we have
\begin{equation}\label{kss}
\forall\,f,g\in L^p(\mathbb{R}^d),\ f(x)g(-i\nabla)\in\mathfrak{S}_{p}(\hl)\text{\ and\ }\ns{p}{f(x)g(-i\nabla)}\le (2\pi)^{-d/p}\nlp{p}{f}\nlp{p}{g}.
\end{equation}
In particular the \emph{density} $\rho_Q$ of $Q$, given by the formula
\[
\forall\,x\in \RR,\ \rho_Q(x):=\ttr_{\mathbb{C}^4}\big(Q(x,x)\big)
\] 
is well defined. In \cite{ptf} Hainzl \emph{et al.} prove that its Coulomb energy is finite $D(\rho_Q,\rho_Q)<+\infty$. By Cauchy-Schwartz inequality, $D(\nu,\rho_Q)$ is defined.

\noindent -- By Kato's inequality
\begin{equation}\label{kato}
\dfrac{1}{|\cdot|}\le \dfrac{\pi}{2}|\nabla|,
\end{equation}
the exchange term is well-defined: this implies that $\nqq{Q}^2\le \tfrac{\pi}{2}\ttr\big(|\nabla|Q^*Q \big)$.

\noindent -- Furthermore the following holds: if $\alpha < \tfrac{4}{\pi}$, then the BDF energy is bounded from below on $\mathcal{K}$ \cite{stab,Sc,at}. Here we assume it is the case.


\medskip

\noindent 2. For $Q\in\mathcal{K}$, its charge is its $\PP$-trace: $q=\ttr_{\PP}(Q)$. So we define charge sectors sets:
\[
\forall\,q\in\RR,\ \mathcal{K}^q:=\big\{Q\in\mathcal{K},\ \ttr(Q)=q\big\}.
\]
A minimizer of $\mathcal{E}^\nu_{\text{BDF}}$ over $\mathcal{K}$ is interpreted as the polarized vacuum in the presence of $\nu$ while minimizer over charge sector $N\in\mathbb{N}$ is interpretreted as the ground state of $N$ electrons in the presence of $\nu$.
We define the energy functional $E^\nu_{\text{BDF}}$:
\begin{equation}
 \forall\,q\in\RR,\ E^\nu_{\text{BDF}}(q):=\inf\big\{\mathcal{E}^\nu_{\text{BDF}}(Q),\ Q\in\mathcal{K}^q\big\}.
\end{equation}

We also write:
\begin{equation}\label{koc}
\mathcal{K}^0_{\mathscr{C}}:=\{ Q\in\mathcal{K},\ \text{Tr}_{\PP}(Q)=0,\ -\Cha Q \Cha =Q\}.
\end{equation}
Lemma \ref{weakclosed} states that this set is sequentially weakly-$*$ closed in $\mathfrak{S}_1^{\PP}(\hl)$.

\begin{notation}
 For an operator $Q\in\mathfrak{S}_2(\hl)$, we write $R_Q$ the operator given by the integral kernel:
 \[
  R_Q(x,y):=\frac{Q(x,y)}{|x-y|}.
 \]

\end{notation}

\subsection{Structure of manifold}

We define
\[
 \mathscr{V}=\big\{P-\PP,\ P^*=P^2=P\in\mathcal{B}(\hl),\ \ttr_{\PP}\big( P-\PP\big)=0\big\}\subset \mathfrak{S}_2(\hl). 
\]
Up to adding $\PP$, we deal with
\[
 \mathscr{M}:=\PP+\mathscr{V}=\big\{P,\ P^*=P^2=P,\ \ttr_{\PP}\big( P-\PP\big)=0\big\}.
\]
From a geometrical point of view, we recall that these sets are Hilbert manifolds: $\mathscr{V}$ lives in the Hilbert space $\mathfrak{S}_2(\hl)$ and $\mathscr{M}$ lives in the affine space $\PP+\mathfrak{S}_2(\hl).$

\begin{proposition}\label{manim}
The set $\mathscr{M}$ is a Hilbert manifold and for all $P\in\mathscr{M}$,
\begin{equation}
 \mathrm{T}_P \mathscr{M}=\{ [A,P],\,A\in\mathcal{B}(\hl),\ A^*=-A\text{\ and\ }PA(1-P)\in\mathfrak{S}_2(\hl)\}.
\end{equation}
Writing
\begin{equation}
 \mathfrak{m}_P:=\{ A\in\mathcal{B}(\hl),\ A^*=-A,\ PAP=(1-P)A(1-P)=0\text{\ and\ }PA(1-P)\in\mathfrak{S}_2(\hl)\},
\end{equation}
any $P_1\in\mathscr{M}$ can be written as $P_1=e^A P e^{-A}$ where $A\in\mathfrak{m}_P$.
\end{proposition}

The BDF energy $\mathcal{E}_{\text{BDF}}^\nu$ is a differentiable function in $\mathfrak{S}_1^{\PP}(\hl)$ with: 
\begin{equation}\label{eqdebdf}
 \left\{ \begin{array}{l}
        \forall\, Q,\delta Q\in\mathfrak{S}_1^{\PP}(\hl),\ \text{d}\mathcal{E}_{\text{BDF}}^\nu(Q)\cdot \delta Q=\text{Tr}_{\PP}\big(D_{Q,\nu}\delta Q\big).\\
        D_{Q,\nu}:=\D+\alpha \big((\rho_Q-\nu)*\frac{1}{|\cdot|}-R_Q\big).
\end{array}\right.
\end{equation}
We may rewrite \eqref{eqdebdf} as follows:
\begin{equation}
\forall\, Q,\delta Q\in\mathfrak{S}_1^{\PP}(\hl),\ \text{d}\mathcal{E}_{\text{BDF}}^\nu(Q)\cdot \delta Q=\text{Tr}_{\PP}\big(\Pi_\Lambda D_{Q,\nu}\Pi_\Lambda \delta Q\big)
\end{equation}
\begin{notation}
 In the case $\nu=0$ we write $D_Q:=D_{Q,0}$.
\end{notation}
\begin{proposition}\label{gragra}
Let $(P,v)$ be in the tangent bundle $\mathrm{T}\mathscr{M}$ and $Q=P-\PP$. Then $[[\Pi_\Lambda D_Q \Pi_\Lambda,P],P]$ is a Hilbert-Schmidt operator in $\mathrm{T}_P\mathscr{M}$ and:
 \begin{equation}\label{difftan}
\mathrm{d}\mathcal{E}_{\text{BDF}}^0(P)\cdot v=\mathrm{Tr}\Big(\big[\big[\Pi_\Lambda D_{Q}\Pi_\Lambda ,P\big],P\big]v\Big).
\end{equation}
\end{proposition}
\begin{remark}
 The operator $[[\Pi_\La D_Q\Pi_\La,P],P]$ is the "projection" of $\Pi_\La D_Q \Pi_\La$ onto $\text{T}_P\mathscr{M}$. It properly defines a vector in the tangent plane which is exactly the \emph{gradient} of $\mathcal{E}_{\text{BDF}}^0$ at the point $P$.
 \begin{equation}\label{defgradient}
  \forall\,P\in\mathscr{M},\ \nabla \mathcal{E}_{\text{BDF}}^0(P)=\big[\big[\Pi_\La D_Q \Pi_\La,P\big],P\big].
 \end{equation}
\end{remark}

We recall $\mathscr{M}_{\mathscr{C}}$ is the set of $\Cha$-symmetric states \eqref{def_mc}.

\begin{proposition}\label{manicsym}
The set $\mathscr{M}_{\mathscr{C}}$ is a \emph{submanifold} of $\mathscr{M}$, which is \emph{invariant} under the flow of $\mathcal{E}_{\text{BDF}}^0$. 
For any $P\in \mathscr{M}_{\mathscr{C}}$, writing
\begin{equation}
 \mathfrak{m}^{\mathscr{C}}_P=\{a\in \mathfrak{m}_P,\ \Cha a \Cha=a\},
\end{equation}
we have
\begin{equation}\label{tangentc}
 \mathrm{T}_P \mathscr{M}_{\mathscr{C}}=\{[a,P],\ a\in \mathfrak{m}_P^{\mathscr{C}}\}=\{v\in\mathrm{T}_P \mathscr{M},\ -\Cha v \Cha=v\}.
\end{equation}
Furthermore, for any $P\in\mathscr{M}_{\mathscr{C}}$ we have
\begin{equation}\label{vanishdens}
\rho_{P-\PP}=0.
\end{equation}
\end{proposition}

\begin{proposition}\label{conn}
 The set $\mathscr{M}_{\mathscr{C}}$ has two connected components $\mathscr{E}_1$ and $\mathscr{E}_{-1}$:
 \begin{equation}
 \forall\,P\in \mathscr{M}_{\mathscr{C}},\ P\in \mathscr{E}_{1}\iff\mathrm{Dim}\,\mathrm{Ran}\,P\cap\mathrm{Ran}\,\PPP\equiv 0[2].
 \end{equation}
 In particular, $\mathscr{E}_1$ contains $\PP$ and $\mathscr{E}_{-1}$ contains any $\PP+\ket{\psi}\bra{\psi}-\ket{\mathrm{C}\psi}\bra{\mathrm{C}\psi}$ where $\psi\in\mathrm{Ran}\,\PPP$.
\end{proposition}

\medskip

\noindent We end this section by stating technical results needed to prove Propositions \ref{manim},\ref{manicsym} and \ref{conn}.

\subsection{Form of trial states}

\begin{theorem}[Form of trial states]\label{structure}
Let $P_1,P_0$ be in $\mathscr{N}$ and $Q=P_1-P_0$. Then there exist $M_+,M_-\in\mathbb{Z}_+$ such that there exist two orthonormal families 
\[
\begin{array}{ll}
(a_1,\ldots,a_{M_+})\cup(e_i)_{i\in\mathbb{N}}& \mathrm{in}\ \mathrm{Ran}\,\PPP,\\
(a_{-1},\ldots,a_{-M_+})\cup(e_{-i})_{i\in\mathbb{N}}&\mathrm{in}\ \mathrm{Ran}\,\PP,
\end{array}
\]
and a nonincreasing sequence $(\la_i)_{i\in\mathbb{N}}\in\ell^2$ satisfying the following properties.
\begin{enumerate}
\item The $a_i$'s are eigenvectors for $Q$ with eigenvalue $1$ (resp. $-1$) if $i>0$ (resp. $i<0$).
\item For each $i\in\mathbb{N}$ the plane $\Pi_i:=\text{Span}(e_i,e_{-i})$ is spanned by two eigenvectors $f_i$ and $f_{-i}$ for $Q$ with eigenvalues $\la_i$ and $-\la_i$.
\item The plane $\Pi_i$ is also spanned by two orthogonal vectors $v_i$ in $\text{Ran}(1-P)$ and $v_{-i}$ in $\text{Ran}(P)$. Moreover $\la_i=\sin(\theta_i)$ where $\theta_i\in (0,\tfrac{\pi}{2})$ is the angle between the two lines $\mathbb{C}v_i$ and $\mathbb{C}e_i$.
\item There holds: \[
Q=\ssum_{i=1}^{M_+}\ket{a_i}\bra{a_i}-\ssum_{i=1}^{M_-}\ket{a_{-i}}\bra{a_{-i}}+\ssum_{j\in \mathbb{N}}\la_j(\ket{f_j}\bra{f_j}-\ket{f_{-j}}\bra{f_{-j}}).
\]
\end{enumerate}
\end{theorem}

\begin{remark}
We have
 \begin{equation}\label{++--}
 \begin{array}{l}
  Q^{++}=\ssum_{i=1}^{M_+}\ket{a_i}\bra{a_i}+\ssum_{j\in\mathbb{N}}\sin(\theta_j)^2\ket{e_j}\bra{e_j},\\
  Q^{--}=-\ssum_{i=1}^{M_-}\ket{a_{-i}}\bra{a_{-i}}-\ssum_{j\in\mathbb{N}}\sin(\theta_j)^2\ket{e_{-j}}\bra{e_{-j}}.
 \end{array}
\end{equation}
\end{remark}

Thanks to Theorem \ref{structure}, it is possible to characterize $\Cha$-symmetric states.
\begin{proposition}\label{chasym}
Let $\g=P-\PP$ be in $\mathscr{M}_{\mathscr{C}}$. For $-1\le \mu\le 1$ and $A=\in\{\g,\g^2\}$, we write 
\[E^A_\mu=\text{Ker}(A-\mu).\]
Then for any $\mu\in\sigma(\g)$, we have $\Cha E^\g_\mu=E^\g_{-\mu}$. Moreover for $|\mu|<1$: if we decompose $E^\g_{\mu}\oplus E^\g_{-\mu}$ into a sum of planes $\Pi$ as in Theorem \ref{structure}, then each $\Pi$ is \emph{not} $\Cha$-invariant and $\mathrm{dim}\,E^\g_{\mu}$ is even. Equivalently $\mathrm{dim}\,E^{\g^2}_{\mu^2}$ is divisible by $4$.

Moreover there exists a decomposition 
\[
E^{\g^2}_{\mu^2}=\underset{1\le j\le \tfrac{N}{2}}{\overset{\perp}{\oplus}}V_{\mu,j}\text{\ and\ }V_{\mu,j}=\Pi^a_{\mu,j}\overset{\perp}{\oplus}\Cha \Pi^a_{\mu,j}
\]
where the $\Pi^a_{\mu,j}$'s and $\Cha \Pi^a_{\mu,j}$'s are spectral planes described in Theorem \ref{structure}.
\end{proposition}





\section{Proof of Theorem \ref{critic}}
\subsection{Strategy and tools of the proof}
\begin{center}
 \textsc{Topologies}
\end{center}

 The upper bound in \eqref{lowup} comes from minimization over $\Cha$-symmetric state of form \eqref{imagine}.

We prove the existence of the minimizer over $\mathscr{E}_{-1}$ by using a lemma of Borwein and Preiss \cite{borw,at}, a smooth generalization of Ekeland's Lemma \cite{ek}: we study the behaviour of a specific minimizing sequence $(P_n)_n$ or equivalently $(P_n-\PP=:Q_n)_n$. 

Each element of the sequence satisfies an equation close to the one satisfied by a real minimizer and we show this equation remains in some weak limit.



\begin{remark}\label{topo}
 We recall different topologies over bounded operators, besides the norm topology $\nb{\cdot}$ \cite{ReedSim}.
 \begin{enumerate}
  \item The so-called \emph{strong topology}, the weakest topology $\mathcal{T}_s$ such that for any $f\in\hl$, the map
  \[
   \begin{array}{rcl}
    \mathcal{B}(\hl)&\longrightarrow&\hl\\
    A&\mapsto& Af
   \end{array}
  \]
is continuous.
\item The so-called \emph{weak operator topology}, the weakest topology $\mathcal{T}_{w.o.}$ such that for any $f,g\in\hl$, the map
  \[
   \begin{array}{rcl}
    \mathcal{B}(\hl)&\longrightarrow&\mathbb{C}\\
    A&\mapsto& \psh{A f}{g}
   \end{array}
  \]
  is continuous.
 \end{enumerate}
 We can also endow $\mathfrak{S}_1^{\PP}$ with its weak-$*$ topology, the weakest topology such that the following maps are continuous:
 \[
   \begin{array}{|l}
   \begin{array}{rcl}
    \mathfrak{S}_1^{\PP}&\longrightarrow&\mathbb{C}\\
    Q&\mapsto& \ttr\big(A_0(Q^{++}+Q^{--})+A_2(Q^{+-}+Q^{-+})\big)
   \end{array}\\
  \forall\,(A_0,A_2)\in\mathrm{Comp}(\hl)\times \mathfrak{S}_2(\hl).
  \end{array}
 \]
We emphasize that the weak-$*$ topology is different from the weak topology (where $\mathrm{Comp}(\hl)$ must be replaced by $\mathcal{B}(\hl)$).
\end{remark}

The following Lemma is important in our proof.
\begin{lemma}\label{weakclosed}
 The set $\mathcal{K}^0_{\mathscr{C}}$ (defined in \eqref{koc}) is weakly-$*$ sequentially closed in $\mathfrak{S}_1^{\PP}(\hl)$.
 
\end{lemma}

 We prove this Lemma at the end of this Subsection.
\medskip

\begin{center}
 \textsc{Borwein and Preiss Lemma}
\end{center}
We recall this Theorem as stated in \cite{at}:

\begin{theorem}\label{bp_lemma}
 Let $\mathcal{M}$ be a closed subset of a Hilbert space $\mathcal{H}$, and $F:\mathcal{M}\to (-\infty,+\infty]$ be a lower semi-continuous function that is bounded from below and not identical to $+\infty$.
 For all $\eps>0$ and all $u\in \mathcal{M}$ such that $F(u)<\inf_{\mathcal{M}}+\eps^2$, there exist $v\in\mathcal{M}$ and $w\in\ov{\mathrm{Conv}(\mathcal{M})}$ such that
 
 \begin{enumerate}
  \item $F(v)< \inf_{\mathcal{M}}+\eps^2$,
  \item $|| u-v||_{\mathcal{H}}<\sqrt{\eps}$ and $|| v-w||_{\mathcal{H}}<\sqrt{\eps}$,
  \item $F(v)+\eps || v-w||_{\mathcal{H}}^2=\min\big\{F(z)+\eps || z-w||_{\mathcal{H}}^2,\ z\in\mathcal{M}\big\}.$
 \end{enumerate}
\end{theorem}

Here we apply this Theorem with $\mathcal{H}=\mathfrak{S}_2(\hl)$, $\mathcal{M}=\mathscr{E}_{-1}-\PP$ and $F=\mathcal{E}^0_{\mathrm{BDF}}$.

The BDF energy is continuous in the $\mathfrak{S}_1^{\PP}$-norm topology, thus its restriction over $\mathscr{V}$ is continuous in the $\mathfrak{S}_2(\hl)$-norm topology.

This subspace $\mathcal{H}$ is closed in the Hilbert-Schmidt norm topology because $\mathscr{V}=\mathscr{M}_{\mathscr{C}}$ is closed in $\mathfrak{S}_2(\hl)$ and $\mathscr{E}_{-1}-\PP$ is closed in $\mathscr{V}$.

Moreover, we have
\[
 \ov{\text{Conv}(\mathscr{E}_{-1}-\PP)}^{\mathfrak{S}_2}\subset \mathcal{K}_{\mathscr{C}}^0.
\]

\medskip

For every $\eta>0$, we get a projector $P_\eta\in\mathscr{E}_{-1}$ and $A_\eta\in \mathcal{K}_{\mathscr{C}}^0$ such that $P_{\eta}$ that minimizes the functional
\[
 F_\eta: P\in\mathscr{E}_{-1}\mapsto \mathcal{E}_{\text{BDF}}^0(P-\PP)+\eps \ns{2}{P-\PP-A_\eta}^2.
\]
We write
\begin{equation}\label{almost}
 Q_\eta:= P_\eta -\PP,\ \G_\eta:=Q_\eta -A_\eta,\ \wt{D}_{Q_\eta}:=\Pi_\La \big(\D-\alpha R_{Q_\eta}+2\eta \G_\eta\big)\Pi_\La.
\end{equation}
Studying its differential on $\text{T}_{P_\eta} \mathscr{M}_{\mathscr{C}}$, we get that
\begin{equation}\label{eq_almost}
 \big[\wt{D}_{Q_\eta}, P_\eta\big]=0.
\end{equation}
In particular, by functional calculus, we get that
\begin{equation}\label{pimoins}
 \big[\boldsymbol{\pi}_-^{\eta},P_\eta\big]=0,\ \boldsymbol{\pi}_{\eta}^-:=\chi_{(-\infty,0)}(\wt{D}_{Q_\eta}).
\end{equation}
We also write
\begin{equation}\label{piplus}
 \boldsymbol{\pi}_{\eta}^+:=\chi_{(0,+\infty)}(\wt{D}_{Q_\eta})=\Pi_\La-\boldsymbol{\pi}_{\eta}^-.
\end{equation}
We can decompose $\hl$ as follows (here R means $\text{Ran}$):
\begin{equation}\label{decomp_hl}
 \hl=\text{R}(P_\eta)\cap \text{R}(\pvt_{\eta}^-)\overset{\perp}{\oplus}\text{R}(P_\eta)\cap \text{R}(\pvt_{\eta}^+)\overset{\perp}{\oplus}\text{R}(\Pi_\La-P_\eta)\cap \text{R}(\pvt_{\eta}^-)\overset{\perp}{\oplus}\text{R}(\Pi_\La-P_\eta)\cap \text{R}(\pvt_{\eta}^+).
\end{equation}

We will prove
\begin{enumerate}
 \item $\text{Ran}\,P\cap \text{Ran}\,\pvt_{\eta}^+$ has dimension $1$, spanned by a unitary $\psi_\eta\in\hl$.
 \item As $\eta$ tends to $0$, up to translation and a subsequence, $\psi_\eta\rightharpoonup \psi_e\neq 0$, $Q_\eta\rightharpoonup \ov{Q}$.
       There holds $\ov{Q}+\PP\in \mathscr{E}_{-1}$, $\psi_e$ is a unitary eigenvector of $\Pi_\La D_{\ov{Q}} \Pi_\La$ and
       \[
        \ov{Q}+\PP=\chi_{(-\infty,0)}\big(\Pi_\La D_{\ov{Q}} \Pi_\La \big)+\ket{\psi_e}\bra{\psi_e}-\ket{\Cha \psi_e}\bra{\Cha \psi_e}.
       \]
\end{enumerate}

In the following part we write the spectral decomposition of trial states and prove Lemma \ref{weakclosed}.

\medskip

\begin{center}
 \textsc{Spectral decomposition}
\end{center}

Let $(Q_n)_n$ be any minimizing sequence for $E_{1,1}$. We consider the spectral decomposition of the trial states $Q_n$: thanks to the upper bound, $\text{Dim}\,\text{Ker}(Q_n-1)=1$, as shown in Subsection \ref{subscritic}.

There exist a \emph{non-increasing} sequence $(\la_{j;n})_{j\in\mathbb{N}}\in\ell^2$ of eigenvalues and an orthonormal basis $\mathbf{B}_n$ of $\text{Ran}\, Q_n$:
\begin{equation}\label{basen}
\mathbf{B}_n:=(\psi_n,\Cha\psi_n)\cup (e_{j;n}^a,e_{j;n}^b,\Cha e_{j;n}^a,\Cha e_{j;n}^b),\ \PP \psi_n=\PP e_{j;n}^{\star}=0,\ \star\in\{a,b\},
\end{equation}
such that the following holds. We omit the index $n$:
\begin{subequations}\label{formtrial}
\begin{equation}\label{formtrial1}
\forall\,j\in\mathbb{N},\ \ e_{-j}^a:=-\Cha e_{j}^b,\ e_{-j}^b:=\Cha e_j^a,
\end{equation}

\begin{equation}\label{formtrial2}
 \begin{array}{rll}
  f_{j}^\star&:=& \sqrt{\tfrac{1-\la_j}{2}} e_{-j}^\star+\sqrt{\tfrac{1+\la_j}{2}}e_{j}^\star,\\
  f_{-j}^\star&:=& -\sqrt{\tfrac{1+\la_j}{2}}e_{-j}^\star+\sqrt{\tfrac{1+\la_j}{2}} e_{j}^\star.
 \end{array}
\end{equation}

\begin{equation}\label{formtrial3}
\left\{\begin{array}{rll}
 Q_n&=&\ket{\psi_n}\bra{\psi_n}-\ket{\Cha \psi_n}\bra{\Cha\psi_n}+\ssum_{j\ge 1}\la_jq_{j;n} \\
 q_{j;n}&=&\ket{f_{j}^a}\bra{f_{j}^a}-\ket{f_{-j}^a}\bra{f_{-j}^a}+\ket{f_{j}^b}\bra{f_{j}^b}-\ket{f_{-j}^b}\bra{f_{-j}^b}.
 \end{array}
 \right.
\end{equation}
\end{subequations}

\begin{remark}\label{diag_extrac}
Thanks to the cut-off, the sequences $(\psi_n)_n$ and $(e_{j;n})_n$ are $H^1$-bounded. 

Up to translation and extraction ($(n_k)_k\in\mathbb{N}^{\mathbb{N}}$ and $(x_{n_k})_k\in(\mathbb{R}^3)^{\mathbb{N}}$), we assume that the weak limit of $(\psi_n)_n$ is non-zero (if it were then there would hold $E_{1,1}=2m$).

We consider the weak limit of each $(e_n)$: by means of a diagonal extraction, we assume that all the $(e_{j,n_{k}}(\cdot -x_{n_k}))_k$ and $(\psi_{j,n_k}(\cdot-x_{n_k}))_k$, converge along the same subsequence $(n_k)_k$. We also assume that
\begin{equation}\label{spec_conv}
 \forall\,j\in\mathbb{N},\ \la_{j,n_k}\to\mu_j,\ (\mu_j)_j\in\ell^2,\ (\mu_j)_j\text{\ non-increasing},
\end{equation}
and that the above convergences also hold in $L^2_{\text{loc}}$ and almost everywhere.
\end{remark}


\medskip

\begin{center}
 \textsc{Proof of Lemma \ref{weakclosed}}
\end{center}


Let $(Q_n)_n$ be a sequence in $\mathcal{K}^0_{\mathscr{C}}$ that converges to $Q\in\mathcal{K}$ in the weak-$*$ topology of $\mathfrak{S_1}^{\PP}$, that is:
\begin{subequations}
\begin{equation*}
 \forall (G_0,G_2)\in\text{Comp}(\hl)\times\mathfrak{S}_2(\hl):
 \end{equation*}
 \begin{equation*}
 \left\{\begin{array}{lcl}
  \text{Tr}(Q_n^{+-}G_2)\underset{n\to+\infty}{\to}\text{Tr}(Q^{+-}G_2)&\text{and}&\text{Tr}(Q_n^{-+}G_2)\underset{n\to+\infty}{\to}\text{Tr}(Q^{-+}G_2),\\
  \text{Tr}(Q_n^{++}G_0)\underset{n\to+\infty}{\to}\text{Tr}(Q^{++}G_0)&\text{and}&\text{Tr}(Q_n^{--}G_0)\underset{n\to+\infty}{\to}\text{Tr}(Q^{--}G_0).
 \end{array}
\right.
\end{equation*}
\end{subequations}
In particular we have $S:=\text{sup}_n\ns{2}{Q_n}<+\infty$ by the uniform boundedness principle. The $\Cha$-symmetry is a weak-$*$ condition: for all $\phi_1,\phi_2\in\hl$:
\[
 \text{Tr}\big(-\Cha Q_n \Cha \ket{\phi_1}\bra{\phi_2}\big)=-\psh{Q_n\Cha \phi_1 }{\Cha \phi_2}
\]
thus $-\Cha Q\Cha=Q$. There remains to prove that $\text{Tr}_{\PP}(Q)=0$. 




We consider the spectral decomposition of $P_n:=\PP+Q_n$. We know that this is compact perturbation of $\PP$, thus its essential spectrum is $\{0,1\}$ and there exist an ONB of $\hl$:
\[
(e_{k;n})_{k=1}^{K_1}\cup(f_{j;n})_{j\in\mathbb{N}}\cup(g_{j;n})_{j\in\mathbb{N}},\ K_1\in\mathbb{Z}_+
\]
and two sequences $(r_{j;n})_j,(s_{j;n})_j$ in $[0,\tfrac{1}{2})$ that tend to $0$, such that
\[
P_n=\dfrac{1}{2}\ssum_{k=1}^{K_1}\ket{e_{k;n}}\bra{e_{k;n}}+\ssum_{j\in\mathbb{N}}\Big\{ r_{j;n}\ket{f_{j;n}}\bra{f_{j;n}}+(1-s_{j;n})\ket{g_{j;n}}\bra{g_{j;n}}\Big\}.
\]
Our aim is to prove we can rewrite $P_n$ as follows:
\begin{equation}\label{aim_weak}
\begin{array}{|l}
P_n=\ov{P}_n+\ov{\g}_n,\\
\ov{\g}_n=\ssum_{j}t_{j;n}\big(\ket{\phi_{j;n}}\bra{\phi_{j;n}}-\ket{\Cha \phi_{j;n}}\bra{\Cha \phi_{j;n}}\big),\\
\ov{P}_n\in \mathscr{M}_{\mathscr{C}},\ 2\sum_{j}t_{j;n}\le \ttr\big(Q_n^{++}-Q_n^{--}\big),\\
(\phi_{j;n})_j\cup(\Cha \phi_{j;n})_j\ \text{orthonormal\ family}.
\end{array}
\end{equation}

Let us assume this point for the moment. Up to extraction, it is clear that the weak limit $\ov{\g}_{\infty}$ of $(\ov{\g}_n)$ has trace $0$: the eventual loss of mass of $(\phi_{j;n})_n$ is compensated by that of $(\Cha \phi_{j;n})_n$: $|\phi_{j;n}(x)|^2=|\Cha \phi_{j;n}(x)|^2$ for all $x\in\RR$. So the weak limit of 
\[t_{j;n}\big(\ket{\phi_{j;n}}\bra{\phi_{j;n}}-\ket{\Cha \phi_{j;n}}\bra{\Cha \phi_{j;n}}\big)
\]
has trace $0$.

The same goes for $\ov{Q}_n:=\ov{P}_n-\PP$. We write $S:=\limsup_n \ttr_{\PP}(\ov{Q}_n)<+\infty$.
We decompose each $\ov{Q}_n$ as in \eqref{formtrial} and take the same notations. We may have $D_n:=\text{Dim}(\ov{Q}_n^2-1)>2$ but the sequence $(D_n)_n$ is bounded by $S$. There is at most $\tfrac{S}{2}$ different $\psi_{j;n}$ in the spectral decomposition of $\ov{Q}_n$ ($j=1,\ldots,\lfloor\tfrac{S}{2}\rfloor$).

We study the weak-limit of the $\psi_{j;n}$'s and the $e_{j;n}^\star$'s: there may be a loss of mass. However from \eqref{++--}, we see that the loss of mass in $\psi_{j;n}$ is compensated by that of $\Cha \psi_{j;n}$, and that of $e_{j;n}^\star$ is compensated by that of $\Cha e_{j;n}^\star$. 

The subscript $\infty$ means we take the weak limit. If the sequences of eigenvalues $(\la_j)_j\in\ell^2$ weakly converges to $(\mu_j)_j\in\ell^2$, then we get that
\[
 \left\{\begin{array}{l}
  Q^{++}=\ssum_{1\le j\le \lfloor S/2\rfloor}\ket{\psi_{j;\infty}}\bra{\psi_{j;\infty}}+\ssum_{j\in\mathbb{N}}\mu_j^2\Big\{\ket{e_{j,\infty}^a}\bra{e_{j,\infty}^a}+\ket{e_{j,\infty}^b}\bra{e_{j,\infty}^b}\Big\}\\
  Q^{--}=-\ssum_{1\le j\le \lfloor S/2\rfloor}\ket{\psi_{-j;\infty}}\bra{\psi_{-j;\infty}}-\ssum_{j\in\mathbb{N}}\mu_j^2\Big\{\ket{e_{-j,\infty}^a}\bra{e_{-j,\infty}^a}+\ket{e_{-j,\infty}^b}\bra{e_{-j,\infty}^b}\Big\}
 \end{array}\right.
\]
where $|\psi_{j,\infty}|^2=|\psi_{-j,\infty}|^2$ resp. $|e^\star_{j,\infty}|^2=|e^\star_{-j,\infty}|^2$. Thus
\[
 \text{Tr}\big(Q^{++}+Q^{--}\big)=0.
\]

\paragraph{Proof of \eqref{aim_weak}} The condition $-\Cha Q_n \Cha =Q_n$ is equivalent to $\Cha P_n \Cha = \Pi_\La-P_n$, so for any $\mu\in \mathbb{R}$ we have
\[
\Cha \text{Ker}\,\big(P_n-\mu\big)=\text{Ker}\,\big(P_n-(1-\mu)\big).
\]
Up to reindexing the sequences, we can assume that $r_{j;n}=s_{j;n}$ and up to changing the ONB, we can assume that $g_{j;n}=\Cha f_{j;n}$. Let us remark that
\[
\Cha B_n\Cha =B_n\text{\ where\ }B_n:=\dfrac{1}{2}\ssum_{k=1}^{K_0}\ket{e_{k;n}}\bra{e_{k;n}}.
\]
As shown in \cite[Lemma 15, Appendix B]{at}, the condition $Q_n\in\mathfrak{S}_1^{\PP}$ gives
\[
\begin{array}{rcl}
\ttr\big(Q_n^{++}-Q_n^{--}\big)&=& \dfrac{K_1}{2}+\ssum_{j\ge 1}\Big\{r_{j;n}\nlp{2}{\PPP f_{j;n}}^2+(1-r_{j;n})\nlp{2}{\PP f_{j;n}}^2\Big\}\\
   &&+\ssum_{j\ge 1}\Big\{(1-s_{j;n})\nlp{2}{\PPP g_{j;n}}^2+s_{j;n}\nlp{2}{\PP g_{j;n}}^2\Big\},
\end{array}
\]
which implies
\[
\begin{array}{|l}
\dfrac{K_1}{2}+\ssum_{j\ge 1}(r_{j;n}+s_{j;n})\le \ttr_{\PP}(Q_n).
\end{array}
\]
In particular we can write
\[
\begin{array}{|l}
P_n=\ov{P}_n+\g_n+B_n,\\
\g_n=\ssum_{j\ge 1}r_{j:n}\big(\ket{f_{j;n}}\bra{f_{j;n}}-\ket{\Cha f_{j;n}}\bra{\Cha f_{j;n}} \big),\\
P'_n=\ssum_{j\ge 1} \ket{\Cha f_{j;n}}\bra{\Cha f_{j;n}}.
\end{array}
\]
Both $\g_n$ and $B_n$ are trace-class, thus $P'_n-\PP\in\mathfrak{S}_1^{\PP}$. We know that $\ttr_{\PP}(P'_n-\PP)$ is an integer \cite{ptf}, this gives
\[
\frac{K_1}{2}=K_0\in\mathbb{N}.
\]
Let us prove that we can decompose $\text{Ran}\,B_n$ as follows:
\begin{equation}\label{induc}
\text{Ran}\,B_n=F_n\overset{\perp}{\oplus}\Cha F_n,\ \text{Dim}\,F_n=K_0.
\end{equation}
This ends the proof: we have
\[
B_n=\text{Proj}(\Cha F_n)+\frac{1}{2}\big(\text{Proj}(F_n)-\text{Proj}(\Cha F_n)\big)
\]
where $\text{Proj}(E)$ is the orthogonal projection onto $E$. We choose then
\[
\begin{array}{|l}
\ov{P}_n:=P'_n+\text{Proj}(\Cha F_n),\\
\ov{\g}_n:=\g_n+\frac{1}{2}\big(\text{Proj}(F_n)-\text{Proj}(\Cha F_n)\big).
\end{array}
\]

Let $\phi\in \text{Ran}\,B_n$ with $\Cha \phi\notin\mathbb{C}\phi$. Else, we take $\phi\perp \phi'$ with
\[
\Cha \phi=e^{i\theta} \phi,\ \Cha \phi'=e^{i\theta'}\phi',\ \theta,\theta'\in\mathbb{R}.
\]
Up to considering $e^{i\theta/2}\phi$ and $e^{i\theta'/2}\phi'$ we may assume that $\Cha \phi=\phi$, $\Cha \phi'=\phi'$.
Then writing
\[
\phi_{\pm}:=\dfrac{1}{\sqrt{2}}\big( \phi\pm i \phi'\big)
\]
we have $\psh{\Cha \phi_+}{\phi_+}=0$, which is absurd.

Let us consider $\text{Span}(\phi,\Cha \phi)$ and assume $\nlp{2}{\phi}=1$. Thus $z=\psh{\Cha \phi}{\phi}=-re^{i\theta}$ with $0\le r\le 1$.
There exist $a,b\in\mathbb{C}$ such that
\[
\psh{\Cha (a \phi+b \Cha \phi)}{a\phi +b\Cha \phi}=0.
\]
If $r=0$ we take $a=1$ and $b=0$, else it suffices to take $a=r_0e^{-i\theta/2}$ and $b=r_1e^{i\theta/2}$ where $r_0,r_1>0$ are any number that satisfies
\[
\frac{r_0}{r_1}+\frac{r_1}{r_0}=\frac{2}{r}.
\]
This is possible because as $0<r\le 1$ we have $\tfrac{2}{r}\ge 2$. By an easy induction, we can write $\text{Ran}\,B_n$ as in \eqref{induc}.

\hfill{\small$\Box$}


\subsection{Upper and lower bounds of $E_{1,1}$}\label{subscritic}
\begin{center}
 \textsc{Upper bound}
\end{center}
 We consider trial states of the following form:
\[
 Q=\ket{\psi}\bra{\psi}-\ket{\Cha\psi}\bra{\Cha\psi},\ \nlp{2}{\psi}=1\text{\ and\ }\PP\psi=0.
\]
The set of these states is written $\mathscr{E}_{-1}^0$. We will prove that the energy of a particular $Q$ gives the upper bound. For such a $Q$, the BDF energy is simply:
\begin{equation}\label{form_no_pol}
 2\psh{|\D|\psi}{\psi}-\frac{\alpha}{2}\diint\frac{|\psi\wedge\Cha\psi(x,y)|^2}{|x-y|}dxdy.
\end{equation}


Following \cite{sok}, we take $\phi_{\mathrm{CP}}\in L^2(\RR,\mathbb{C})$ the unique positive radial minimizer of the Choquard-Pekar energy. \emph{We know that this minimizer is in the Schwartz class} (here we just need it to be in $H^2$). We form the spinor:
\[
 \phi:=\begin{pmatrix}\phi_{\mathrm{CP}}&0& 0&0 \end{pmatrix}^{\text{T}},
\]
and scale $\phi$ by a constant $\la^{-1}\sim \alpha$ to be chosen later:
\[
 \phi_\la(x):=\la^{-3/2}\phi(x/\la).
\]
We define $\psi_\la:=\Pi_\La \phi_\la$ and write:
\begin{equation}
 \psi_+:=\frac{\PPP \psi_\la}{\nlp{2}{\PPP \psi_\la}}\text{\ and\ }\psi_-:=\frac{\PP\Cha \psi_\la}{\nlp{2}{\PP\Cha \psi_\la}}=\Cha \psi_+.
\end{equation}
Let us compute the energy of 
\begin{equation}
Q_0:=\ket{\psi_+}\bra{\psi_+}-\ket{\psi_-}\bra{\psi_-}. 
\end{equation}

We have:
\begin{align*}
 \nlp{2}{\PPP \psi_\la}^2&=\underset{B(0,\La)}{\dint}|\wh{\psi_\la}(p)|^2\frac{g_0(p)}{2}\big(1+\tfrac{1}{\ed{p}}\big)dp,\\
                         &=\underset{B(0,\La)}{\dint}|\wh{\psi_\la}(p)|^2g_0(p)\big(1-\tfrac{g_1(p)^2}{4g_0(p)^2}\big)dp+\mathcal{O}(\la^{-4}),\\
                         &=\underset{B(0,\La)}{\dint}|\wh{\psi_\la}(p)|^2\big(m-\tfrac{g_1'(0)^2}{4m}\big)dp+\mathcal{O}((\alpha+\la^{-2})\la^{-2}),\\
                         &=1-\frac{g'_1(0)^2}{4\la^2m}\nlp{2}{\phi_{\mathrm{CP}}}^2+\mathcal{O}((\alpha+\la^{-2})\la^{-2}).
\end{align*}
Similarly the following holds:
\begin{align*}
 \psh{|\D|\PPP \psi_\la}{\psi_\la}&=\underset{B(0,\La)}{\dint}\ed{p}\psh{\wh{\PPP}(p)\wh{\psi}_\la(p)}{\wh{\psi}_\la(p)}_{\mathbb{R}^3}dp\\
                                  &=\underset{B(0,\La)}{\dint}|\wh{\psi_\la}(p)|^2\frac{1}{2}\big(g_0(p)+\ed{p}\big)dp\\
                                  &=m+\frac{g'_1(0)^2}{4\la^2m}\nlp{2}{\phi_{\mathrm{CP}}}^2+\mathcal{O}((\alpha+\la^{-2})\la^{-2}).
\end{align*}
Then we estimate:
\begin{align*}
\diint\frac{|\psi_+\wedge\psi_-(x,y)|^2}{|x-y|}dxdy&=2\Big\{D\big(|\psi_+|^2,|\psi_-|^2\big)-D\big(\psi_+^*\psi_-,\psi_+^*\psi_-\big)\Big\}\\
                                                   &=2\Big\{\tfrac{1}{\la}D\big(|\phi_{\mathrm{CP}}|^2,|\phi_{\mathrm{CP}}|^2\big)+\mathcal{O}(\la^{-2})-D\big(\psi_+^*\psi_-,\psi_+^*\psi_-\big)\Big\}\\
                                                   &=2\Big\{\tfrac{1}{\la}D\big(|\phi_{\mathrm{CP}}|^2,|\phi_{\mathrm{CP}}|^2\big)+\mathcal{O}(\la^{-2})\Big\}.
\end{align*}
Indeed we have:
\[
\begin{array}{cll}
 \nlp{1}{\psi_+^*\psi_-}&\apprle& \nlp{2}{\nabla \psi_\la}\nlp{2}{\psi_\la}=\mathcal{O}(\la^{-1}).\\
 |\psi_+^*\psi_-|*\tfrac{1}{|\cdot|}&\le& |\psi_+|^2*\tfrac{1}{|\cdot|}\le \tfrac{\pi}{2}\psh{|\nabla|\psi_+}{\psi_+}\\
                                  &=& \mathcal{O}(\la^{-1}).
\end{array}
\]
Thus we get that:
\begin{equation}
 \mathcal{E}_{\mathrm{BDF}}^0(Q_0)=2m+\frac{g'_1(0)^2}{\la^2m}\nlp{2}{\nabla \phi_{\mathrm{CP}}}^2-\frac{\alpha}{\la}D\big(|\phi_{\mathrm{CP}}|^2,|\phi_{\mathrm{CP}}|^2\big)+\mathcal{O}((\alpha+\la^{-2})\la^{-2}).
\end{equation}
If we \emph{choose}
\begin{equation}
 \frac{1}{\la}:=\frac{\alpha m}{g'_1(0)^2}
\end{equation}
we get the following upper bound:
\begin{equation}
 E_{1,1}\le\mathcal{E}_{\mathrm{BDF}}^0(Q_0)=2m+\alpha^2\frac{m}{g'_1(0)^2}E_{\mathrm{CP}}+\mathcal{O}(\alpha^3).
\end{equation}

\medskip
\begin{center}
 \textsc{A priori lower bound}
\end{center}


Let $Q\in \mathscr{M}-\PP$ be an approximate minimizer such that
\[
 \mathcal{E}^0_{\text{BDF}}(Q)<E_{1,1}+\alpha^2\frac{m}{2g'_1(0)^2}|E_{\mathrm{CP}}|<2m.
\]
Our aim is to prove the following
\begin{equation}\label{a_priori}
\left\{
	\begin{array}{rcl}
		E_{1,1}-2m&\ge&-K\alpha^2,\\
		\ttr\big(|\nabla|Q^2\big)&\le&K\alpha.
	\end{array}
\right.
\end{equation}

We have
\[
 \Big(1-\alpha\frac{\pi}{4}\Big)\text{Tr}(|\D|Q^2)\le\mathcal{E}^0_{\text{BDF}}<2m\text{\ so\ }\ns{2}{Q}^2<\frac{2m}{1-\alpha\frac{\pi}{4}}<3.
\]
However $\ns{2}{Q}^2\ge \text{Dim}\,\text{Ker}(Q^2-1)=2\text{Dim}\,\text{Ker}(Q-1)$, thus $Q$ has the form written in \eqref{formtrial}; in particular we have:
\[
 Q=\ket{\psi}\bra{\psi}-\ket{\Cha\psi}\bra{\Cha \psi}+\g,\ \psi\in\text{Ran}(\PPP),\ \psi_+:=\psi,\psi_-:=\Cha\psi\in\text{Ker}\,\g.
\]
Let us remark that $\g+\PP\in\mathscr{M}$. The energy of $Q$ is:
\begin{equation}\label{estimtrial}
 \mathcal{E}_{\text{BDF}}^0(Q)=\mathcal{E}_{\text{BDF}}^0(\g)+2\psh{|\D|\psi}{\psi}-\frac{\alpha}{2}\diint\frac{|\psi\wedge\Cha\psi(x,y)|^2}{|x-y|}dxdy-\alpha\ssum_{\eps\in\{+,-\}}\big(\psh{\psi_\eps R_\g}{\psi_\eps}\big).
\end{equation}
We substract $2m$: as $g'_0(0)=0$ and $\nlp{\infty}{g_0''}\le K\alpha$ \cite[Appendix A]{sok}, we have
\[
 |g_0(p)-m|\le p^2\dint_0^1|g_0''(tp)|(1-t)dt\le K\alpha p^2,
\]
thus:
\begin{align*}
 \ed{p}-m&=\frac{g_1(p)^2+(g_0(p)-m)(g_0(p)+m)}{\ed{p}+m}\\
         &\le\frac{g_1(p)^2(1-K\alpha)}{2\ed{p}}.
\end{align*}
Going back to the energy, we have by Cauchy-Schwartz inequality:
\[
 |\psh{\psi_\eps}{R_\g \psi_\eps}|\le \nqq{N[\psi_\eps]}\nqq{\g},\ N[\psi_\eps]:=\ket{\psi_\eps}\bra{\psi_\eps}.
\]
The quantity $\nqq{N[\psi_\eps]}^2$ is simply $D\big(|\psi_\eps|^2,|\psi_\eps|^2\big)$ and we get:
\[
\begin{array}{l}
 (1-K\alpha)\psh{\tfrac{g_1^2(-i\nabla)}{|\D|} \psi}{\psi}+\text{Tr}(|\D|\g^2)\le K_1\alpha^2+2\alpha D\big(|\psi|^2,|\psi|^2\big)+\frac{3\alpha}{2}\nqq{\g}^2,\\
 (1-K\alpha)\psh{\tfrac{g_1^2(-i\nabla)}{|\D|} \psi}{\psi}+\big(1-\frac{3\alpha\pi}{4}\big)\text{Tr}(|\D|\g^2)\le K_1\alpha^2+\alpha \pi\psh{\,|\nabla|\psi}{\psi}.
\end{array}
\]
Now we have:
\begin{equation}
 (1-K\alpha)\frac{p^2}{\ed{p}}\ge 2\alpha |p|\iff p^2\ge 4\alpha^2(1-K\alpha)\ed{p}^2.
\end{equation}
We can take $K=\nlp{\infty}{g_0}$: this inequality holds for
\begin{equation}
 |p|\ge r_0:=\frac{2\alpha\nlp{\infty}{g_0}\sqrt{1-\alpha\nlp{\infty}{g_0}}}{\sqrt{1-4\alpha^2\nlp{\infty}{g'_1}^2(1-\alpha\nlp{\infty}{g_0})}}.
\end{equation}
If we split $\psh{|\nabla \psi|}{\psi}$ at level $|p|=r_0$, we have:
\begin{equation}
 \frac{1-\nlp{\infty}{g_0}\alpha}{2}\psh{\tfrac{g_1^2(-i\nabla)}{|\D|} \psi}{\psi}+\big(1-\frac{3\alpha\pi}{4}\big)\text{Tr}(|\D|\g^2)\le K_1\alpha^2+\alpha r_0\apprle \alpha^2.
\end{equation}
and
\begin{equation}
 \psh{|\nabla|\psi}{\psi}\apprle \alpha.
\end{equation}

Substituting these estimates in \eqref{estimtrial}, we get:
\begin{equation}
 E_{1,1}-2m\ge \mathcal{E}_{\text{BDF}}^0(Q)-2m+\alpha^2\frac{m}{2g'_1(0)^2}E_{\mathrm{CP}}\ge -K\alpha^2.
\end{equation}

\medskip

\begin{center}
 \textsc{Form of a minimizer for $E_{1,1}$}
\end{center}

If a minimizer $\ov{P}\in\mathscr{E}_{-1}$ exists, then it satisfies the following: 
\[
\begin{array}{|l}
\ov{P}=\PP+\ov{Q}=\PP+\ket{\psi_+}\bra{\psi_+}-\ket{\Cha \psi_+}\bra{\Cha \psi_+}+\g\\
\psi_+,\Cha \psi_+\in  \text{Ker}\,\g,\ \PP \psi_+=0.
\end{array}
\]
Moreover the proof of the lower bound ensures that $\ns{2}{\g}\apprle \alpha$. So let $\text{P}^0_{1,1}$ be:
\[
\text{P}^0_{1,1}:=\PP+\ket{\psi_+}\bra{\psi_+}-\ket{\Cha \psi_+}\bra{\Cha \psi_+}.
\]
Then we have $\ns{2}{\text{P}^0_{1,1}-\ov{P}}=\ns{2}{\g}\apprle \alpha$. Using Propositions \ref{manim} and \ref{manicsym}, we write 
\[
\ov{P}=e^A \text{P}^0_{1,1}e^{-A},\ A\in\mathfrak{m}^{\mathscr{C}}_{\text{P}^0_{1,1}}
\]
where there exist $(\theta_j)_j\in\ell^2$ decreasing and $K_0>0$ such that
\[
\begin{array}{l}
\ns{2}{\g}=4\ssum_{j=1}^{+\infty}\sin(\theta_j)^2\le K_0\alpha^2,\text{\ thus}\\
\ns{2}{A}^2=4\ssum_{j=1}^{\infty}\theta_j^2\le\frac{\pi^2}{4}K_0 \alpha^2.
\end{array}
\]

Assuming Theorem \ref{critic}, this proves the description of Theorem \ref{foform}.

\subsection{Existence of a minimizer for $E_{1,1}$}
We consider a family of almost minimizers $(P_{\eta_n})_n$ of type \eqref{almost} where $(\eta_n)_n$ is any decreasing sequence. We assume that $\La^2\alpha^{-2}\eta_n$ is small. We also consider the spectral decomposition \eqref{formtrial} of any $Q_n:=P_{\eta_n}-\PP$. 

For short we write $P_n:=P_{\eta_n}$ and in general replace the subscript $\eta_n$ by $n$.

\medskip

\noindent -- We study weak limits of $(Q_{n})_n$. We recall that $\mathbb{C}\psi_n=\text{Ker}(Q_n-1)$, and
\begin{equation}\label{spec_no}
Q_n=\ket{\psi_n}\bra{\psi_n}-\ket{\Cha \psi_n}\bra{\Cha \psi_n}+\g_n,\ \psi_n,\Cha\psi_n\in\text{Ker}\,\g_n.
\end{equation}

\noindent -- We first prove that there is no vanishing:
\[
\exists A>0,\ \limsup_n \sup_{z\in\RR}\underset{B(z,A)}{\dint}|\psi_{n}(x)|^2dx>0.
\]
Indeed, let us assume this is false. Then for any $A>0$ the following holds:
\[
D\big(|\psi_n|^2,|\psi_n|^2\big)\le  \frac{1}{A}+2\La \Big\{\sup_{z\in\RR}\underset{B(z,A)}{\dint}|\psi_{n}(x)|^2dx\Big\}^{1/2},
\]
where we have used Cauchy-Schwarz inequality and Hardy inequality. In the limit $n\to+\infty$ and then $A\to +\infty$, we have: $\limsup_n D\big(|\psi_{n}|^2,|\psi_{n}|^2\big)=0.$

There holds \emph{a priori} estimates \eqref{a_priori}: using Kato's inequality we would get
\[
\liminf_n \mathcal{E}^0_{\text{BDF}}(Q_n)\ge 2 \liminf_n \psh{|\D|\psi_n}{\psi_n}+\liminf_n\mathcal{E}^0_{\text{BDF}}(\g_n)\ge 2m.
\]

\textbf{Thus, up to translation, we assume that $Q_n\rightharpoonup Q_{\infty}\neq 0$.} 

\noindent -- As the BDF energy is sequential weakly lower continuous \cite{Sc}, we have
\[
E_{1,1}\ge \mathcal{E}_{\text{BDF}}^0(Q_{\infty}).
\]
Our aim is to prove that $Q_{\infty}+\PP\in\mathscr{M}_{\mathscr{C}}$: in other words that $Q_{\infty}$ is a minimizer for $E_{1,1}$.

\noindent -- The spectral decomposition \eqref{spec_no} is not the relevant one: let us prove we can describe $P_n$ in function of the spectral spaces of the "mean-field operator" $\wt{D}_{Q_n}$: the first step is to prove \eqref{spec_yes} below.

We recall that $Q_n$ satisfies Eq. \eqref{eq_almost}, that we have the decomposition \eqref{decomp_hl}.

The following holds:
\begin{align*}
\psh{\wt{D}_{Q_n}\psi_n}{\psi_n}&=\psh{|\D|\psi_n}{\psi_n}+\mathcal{O}\big( \alpha \nlp{2}{\,|\nabla|^{1/2}\psi_n}\ns{2}{\,|\nabla|^{1/2}Q}+\eta_n\ns{2}{\G_{n}}\big)\\
              &=\psh{|\D|\psi_n}{\psi_n}+\mathcal{O}(\alpha^2)\ge m-K\alpha^2.
\end{align*}
Thus $\text{Ran}\,P_n\cap \text{Ran}\,\pvt^{n}_+\neq \{0\}.$ Let us prove this subspace has dimension $1$: we use the minimizing property of $Q_n$. The condition on the first derivative gives \eqref{eq_almost}, what is the condition on the second derivative ? For any $A\in \mathfrak{m}_{P_n}^{\mathscr{C}},$ expanding $e^A P_n e^{-A}-P_n$ in power of $A$, we get that the Hessian $\text{Hess}_{P_n}(F_n)$ of $F_n:=F_{\eta_n}$ at point $P_n$ is
\[
\begin{array}{l}
\forall\,V\in\text{T}_{P_n}\mathscr{M}_{\mathscr{C}},\ A=\big[V,P_n\big],\\
\ \ \ \text{Hess}_{F_n}(P_n;V,V)=\ttr\big(\wt{D}_{Q_n}(A^2P_n-AP_n A)\big)+\eta_n\ns{2}{V}^2-\dfrac{\alpha}{2}\nqq{V}^2.
\end{array}
\]
This Hessian is non-negative. For any unitary $f\perp g$ in $\text{Ran}(\Pi_\La-P_n)$ we choose
\[
A:=\ket{f}\bra{-\Cha g}-\ket{-\Cha g}\bra{f}+\ket{g}\bra{\Cha f}-\ket{\Cha f}\bra{g}\in\mathfrak{m}_{P_n}^{\mathscr{C}}.
\]
As $-\Cha \wt{D}_{Q_n}\Cha=\wt{D}_{Q_n}$, the condition on the Hessian gives
\begin{equation}\label{hess_hess}
2\big(\psh{\wt{D}_{Q_n} f}{f}+\psh{\wt{D}_{Q_n} g}{g}\big)+4\eta_n\ge \dfrac{\alpha}{2}\big| \big|\,\big[A,P_n\big]\big| \big|_{\mathrm{Ex}}^2\ge 0.
\end{equation}
We have $\Cha \psi_n\in \text{Ran}(\Pi_\La-P_n)$ and
\[
\psh{\wt{D}_{Q_n}\Cha \psi_n}{\Cha \psi_n}=-\psh{\wt{D}_{Q_n} \psi_n}{ \psi_n}\le -m+K\alpha^2,
\]
thus necessarily for $n$ large, there is no plane in $\text{Ran}(\Pi_\La-P_n)\cap \text{Ran}(\pvt_-^{n})$, equivalently there is no plane in $\text{Ran}\,P_n\cap \text{Ran}\,\pvt^{n}_+$.

There exists a unitary $\psi_{e;n}\in\hl$ that spans $\text{Ran}\,P_n\cap \text{Ran}\,\pvt^{n}_+$. Equivalently $\psi_{v;n}:=\Cha \psi_{e;n}$ spans the other one.

Thus:
\begin{equation}\label{spec_yes}
P_n=\ket{\psi_{e;n}}\bra{\psi_{e;n}}+\pvt_-^n.
\end{equation}
- We thus write
\begin{equation}
Q_n=\ket{\psi_{e;n}}\bra{\psi_{e;n}}-\ket{\psi_{v;n}}\bra{\psi_{v;n}}+\ov{\g}_n=\ov{N}_n+\ov{\g}_n.
\end{equation}

As $ \text{Ran}\,P_n$ is $\wt{D}_{Q_n}$-invariant and that $\wt{D}_{Q_n}$ is bounded (with a bound that depends on $\La$), necessarily
\[
\wt{D}_{Q_n}\psi_{e;n}=\mu_n\psi_{e;n},\ \mu_n\in \mathbb{R}_+.
\]
The condition on the Hessian enables us to say that
\[
m-\mu_n+2\eta_n\ge 0.
\]

\noindent -- As for $\psi_n$, there is no vanishing for $(\psi_{e,n})_n$ for $\alpha$ sufficiently small: decomposing $\psi_+\in\text{Ran}\,P_n$:
\[
\psi_+=a\psi_{e;n}+\phi,\ \phi\in \text{Ran}\,P_n\cap \text{Ran}\,\pvt_-^n,
\]
we have
\[
|a|^2\ge\frac{1}{\mu}\big(m+\psh{|\wt{D}_{Q_n}|\phi}{\phi}-K(\alpha^2+\eta_n\ns{2}{\G_n}) \big).
\]
Provided that $\mu_n$ is close to $1$, the absence of vanishing for $\psi_n$ implies that of $\psi_{e;n}$.

By Kato's inequality \eqref{kato}:
\begin{align*}
\wt{D}_{Q_n}^2&\ge |\D|\big(1-2\alpha\nb{R_{Q_n}|\D|^{-1}}-4\eta_n \nb{\G_n}\big)|\D|\\
			&\ge |\D|^2\big(1-\alpha  \nqq{Q_n}-4\eta_n \ns{2}{\G_n}\big)
\end{align*}
Thus
\[
\big| \wt{D}_{Q_n}\big|\ge |\D|\big(1-\alpha \nqq{Q_n}-2\eta_n \ns{2}{\G_n}\big)\text{\ and\ }\mu_n\ge 1-K(\alpha^2 +\eta_n \ns{2}{\G_n}).
\]
In the same way we can prove that
\[
|\mu_n-m|\apprle \alpha^2+\eta_n\ns{2}{\G_n}
\]
So
\[
\psi_{e,n}\rightharpoonup\psi_{e}\neq 0.
\]

\noindent -- We decompose $\ov{\g_n}=\pvt_-^n-\PP\in\mathscr{E}_1-\PP$ as in \eqref{formtrial}: using Cauchy's expansion \cite{ptf}, we have
\begin{equation}\label{cauchy}
\pvt_-^n-\PP=\frac{1}{2\pi}\dint_{-\infty}^{+\infty}\frac{d \om}{\D+i\omega}\big(2\eta_n \G_n-\alpha\Pi_\La R_{Q_n}\Pi_\La+2\eta_n \G_n \big)\dfrac{1}{\wt{D}_{Q_n}+i\om}\Pi_\La.
\end{equation}
To justify this equality, we remark that $|\wt{D}_{Q_n}|$ is uniformly bounded from below: the r.h.s. of \eqref{cauchy} is well-defined. Integrating the norm of bounded operator in \eqref{cauchy}, we get that
\[
\nb{\pvt_-^n-\PP}\apprle \alpha \nqq{Q_n}+\eta_n\ns{2}{\G_n}<1.
\]

In fact, we can also expand in power of $Y_n:=-\alpha \Pi_\La R_{Q_n}\Pi_\La +2\eta_n \G_n$:
\begin{equation}\label{cauchy2}
\left\{
	\begin{array}{rcl}
		\pvt_-^n-\PP&=&\ssum_{j\ge 1}\alpha^j M_j[B_n],\\
		M_j[Y_n]&=&-\dfrac{1}{2\pi}\dint_{-\infty}^{+\infty}\frac{ d\om}{\D+i\om}\Big(Y_n\frac{1}{\D+i\om} \Big)^{j}.
	\end{array}
\right.
\end{equation}
We take the Hilbert-Schmidt norm \cite{ptf,sok}: as $\ns{2}{R_{Q_n}\tfrac{1}{|\nabla|^{1/2}}}\apprle \nqq{Q}$, we have
\begin{equation}\label{estim_gn}
\ns{2}{\ov{\g}_n}\apprle \alpha \nqq{Q_n}+\eta_n\ns{2}{\G_n}\apprle \alpha^2.
\end{equation}

We thus write
\[
\begin{array}{rcl}
\ov{\g}_n&=&\ssum_{j\ge 1}\la_{j;n}q_{j;n},
\end{array}
\]
where $q_{j;n}$ has the same form as the one in \eqref{formtrial}. 

\noindent -- Up to a subsequence, we assume all weak convergences as in Remark \eqref{diag_extrac}: the sequence of eigenvalues $(\la_{j;n})_n$ tends to $(\mu_j)_j\in\ell^2$ and each $(e_{j;n}^\star)_n$ (with $\star\in\{a,b\}$) tends to $e_{j;\infty}^\star$, $(\psi_{e;n})_n$ tends to $\psi_{e}$. We also assume that the sequence $(\mu_n)_n$ tends to $\mu$ with $0\le \mu\le m$. For shot we write $\psi_v:=\Cha \psi_e$. 

\medskip

\noindent -- We write $\ov{P}:=Q_{\infty}+\PP$ and $\pvac:=\chi_{(-\infty,0)}(D_{Q_\infty})$.  We will prove that

\begin{enumerate}
	\item $\big[D^{(\La)}_{Q_\infty} ,\ov{P}\big]=0$,
	\item $D_{Q_\infty}\psi_{e}=\mu\psi_e$ and so $\pvac \psi_e=0$. 
	
		Moreover $D_{Q_\infty}\Cha \psi_{e}=-\mu\Cha \psi_e$ and $\psh{\Cha \psi_e}{\psi_e}=0$.
	\item $\pvac=\ov{P}-\ket{\psi_e}\bra{\psi_e}+\ket{\Cha \psi_e}\bra{\Cha \psi_e}$.
\end{enumerate}
\begin{notation}
We write $D^{(\La)}_{Q_\infty}:=\Pi_\La D_{Q_\infty} \Pi_\La$ for short.
\end{notation}

This all comes from the fact that
\begin{equation}
\text{s}-\lim_n R_{Q_n}=R_{Q_\infty}.
\end{equation}
This fact enables us to show
\begin{equation}
\begin{array}{|l}
 R_{Q_n}\psi_{e;n}\rightharpoonup_nR_{Q_\infty}\psi_e\text{\ in\ }L^2,\\
\text{s.\,op.}-\lim_n\big(\pvt_-^n-\PP\big)=\pvac-\PP\text{\ in\ }\mathcal{B}(\hl),\\
\text{w.\,op.}-\lim_n P_n=\pvac-\PP+\ket{\psi_e}\bra{\psi_e}-\ket{ \psi_v}\bra{\psi_v}\text{\ in\ }\mathcal{B}(\hl).
\end{array}
\end{equation}

Indeed for any $f\in\hl$ we have
\begin{align*}
\nlp{2}{R_{Q_n}f-R_{Q_{\infty} }f}^2&=\dint \Big|\dint \frac{(Q_n-Q_{\infty})(x,y)}{|x-y|}f(y)dy\Big|^2dx\\
       &\le \nlp{2}{f}^2\Big(\dfrac{1}{A^2}\ns{2}{Q_n-Q_\infty}^2+4\La^2\underset{B(0,2A)^2}{\diint}|(Q_n-Q_\infty)(x,y)|^2dxdy\Big)\\
       &\ \ \ \ \ +4\La^2\ns{2}{Q_n-Q_\infty}^2\underset{B(0,A)^c}{\dint}|f(y)|^2dy.
\end{align*}
We have just split as follows: for $x\in\RR$ we consider
\[
\RR=B(x,A)^c\sqcup B(x,A)\cap B(0,A)\sqcup B(x,A)\cap B(0,A)^c.
\]
Taking the limsup $n\to+\infty$ we get that
\[
\forall\,A>0,\ \limsup_n\nlp{2}{R_{Q_n}f-R_{Q_{\infty} f}}^2\le 4\limsup_n\ns{2}{Q_n}^2\Big(\frac{\nlp{2}{f}^2}{A^2}+4\La^2\underset{B(0,A)^c}{\dint}|f(y)|^2dy\Big),
\]
taking the limit $A\to+\infty$ we get that
\[
\limsup_n\nlp{2}{R_{Q_n}f-R_{Q_{\infty} f}}^2=0.
\]

In particular for any $f\in \hl$
\[
\psh{R_{Q_n} \psi_{e;n}}{f}=\psh{\psi_{e;n}}{R_{Q_n} f}\underset{n\to+\infty}{\longrightarrow}\psh{\psi_e}{R_{Q_\infty} f}=\psh{R_{Q_\infty}\psi_e}{ f}.
\]
Thus $\wt{D}_{Q_n}\psi_{e;n}\underset{n\to+\infty}{\rightharpoonup}D_{Q_\infty}\psi_e$, and $D_{Q_\infty}\psi_e=\mu\psi_e.$


\noindent -- Let us prove that
\begin{equation}\label{slim_pvac}
\text{s.\,op.}-\lim_n \pvt_-^n=\pvac.
\end{equation}
We have
\[
R_{Q_n}\frac{1}{\D+i\om}f=R[Q_n-Q_{\infty}]\frac{1}{\D+i\om}f+R_{Q_\infty}\frac{1}{\D+i\om} f
\]
and at fixed $\om$ and $f$
\[
R[Q_n-Q_{\infty}]\frac{1}{\D+i\om}f\underset{n\to+\infty}{\longrightarrow} 0\ \text{in}\ L^2.
\]
Generally for $J\ge 1$, we expand $\Big(R_{Q_n}\frac{1}{\D+i\om}\Big)^J$ in power of $R[Q_n-Q_\infty]$ and $Q_\infty$. We get:
\[
\forall\, \om,f,\ \Big(R_{Q_n}\frac{1}{\D+i\om}\Big)^J\underset{n\to+\infty}{\longrightarrow} 0\ \text{in}\ L^2.
\]
Moreover

\begin{align*}
\Big| \Big| \Big(R_{Q_n}\frac{1}{\D+i\om}\Big)^J\Big|\Big|_{L^2}&\le \ed{\om}^{-J/2}\nb{Q_n \tfrac{1}{|\D|^{1/2}}}^J\nlp{2}{f},\\
  &\le \big(\limsup_n \nqq{Q_n} \ed{\om}^{-1/2}\big)^J\nlp{2}{f}.
\end{align*}

By dominated convergence as
\begin{equation}\label{tg}
u_j\nlp{2}{f}:=\dint\frac{d\om}{\ed{\om}^{1+J/2}}\big(\alpha \nqq{Q_n}+\eta_n \ns{2}{\G_n}\big)^J\nlp{2}{f}<+\infty,
\end{equation}
we get
\[
M_j[Y_n]f\underset{n\to+\infty}{\longrightarrow}M_j[\alpha R_{Q_\infty}]f\ \text{in}\ L^2.
\]
To end this argument we remark that the series $\ssum_{j\ge 1}u_j$ is convergent for $\alpha$ and $\eta_n$ sufficiently small: thus we have
\[
\ssum_{j\ge 1}M_j[Y_n] f\underset{n\to+\infty}{\longrightarrow}\ssum_{j\ge 1}M_j[\alpha R_{Q_\infty}] f\text{\ in\ }L^2,
\]
that is \eqref{slim_pvac} holds.

\noindent -- Thanks to \eqref{slim_pvac}, there holds (in the weak operator topology for instance)
\[
Q_\infty=\lim_n Q_n=\ket{\psi_e}\bra{\psi_e}-\ket{ \psi_v}\bra{ \psi_v}+\pvac-\PP,
\]
that is
\begin{equation}
\ov{P}=\ket{\psi_e}\bra{\psi_e}-\ket{ \psi_v}\bra{ \psi_v}+\pvac.
\end{equation}
In the weak operator topology we also have
\[
\text{w.\,op.}-\lim_n \big[\wt{D}_{Q_n},Q_n+\PP\big]=\big[ D^{(\La)}_{Q_\infty}, Q_{\infty}+\PP\big],
\]
by strong convergence of $R_{Q_n}$ to $R_{Q_\infty}$ and norm convergence of $\eta_n\G_n$ to $0$.

\noindent -- There remains to prove that $\nlp{2}{\psi_e}=1$. We assume for the moment that we can uniformly separate the $\mu_n$'s from the remainder of the positive spectrum $\sigma\big( |\wt{D}_{Q_n}|\big)\backslash\{\mu_n\}$. Let us write $a_n$ the bottom of this last set: there exists $\eps>0$ (of order $\alpha^2$ in fact) such that for $n_0$ sufficiently large:
\begin{equation}\label{unif}
\forall n\ge n_0,\ a_n-\mu_n\ge 5\eps.
\end{equation}
In particular, we can draw a small circle in $\mathbb{C}$ that intersects $\mathbb{R}$ only at points $\mu\pm 2\eps$. We write $\mathcal{C}_\eps$ this circle: it has been chosen such that if $|\mu_n-\mu|\le \eps$ (true for $n\ge n_1$ where $n_1\ge n_0$ is sufficiently large), 
\[
\forall n\ge n_0,\ \text{dist}\big(\mu_n;\mathcal{C}_\eps \big)\ge  \eps.
\]
By functional calculus we have
\[
\ket{\psi_{e;n}}\bra{\psi_{e;n}}=\dfrac{1}{2i\pi}\dint_{\mathcal{C}_\eps}\dfrac{dz}{z-\wt{D}_{Q_n}}.
\]
We want to substract $\chi_{(\mu-2\eps,\mu+2\eps)}\big( D^{(\La)}_{Q_\infty}\big)$. If \eqref{unif} is true, then the same holds for the limit $D^{(\La)}_{Q_\infty}$ by strong convergence. Indeed, for any $f\in \text{Ran}\big(\pvac_+ \big)$ (where $\pvac_+:=\Pi_\La-\pvac$) we have
\[
\nlp{2}{\pvt^{n}_+ f-f}\to 0.
\]
For $f_1\perp f_2$ in $\text{Ran}\big(\pvac_+ \big)$, there holds
\begin{align*}
\min_j\big( \frac{1}{\nlp{2}{\pvac^n_+ f_j}^{2}}\big)\psh{\wt{D}_{Q_n} \pvac^n_+ f_1}{f_1}+\psh{\wt{D}_{Q_n} \pvac^n_+ f_2}{f_2}&\ge a_n+\mu_n
\end{align*}
\noindent -- We prove the gap \eqref{unif} for $D^{(\La)}_{Q_\infty}$ by taking the liminf.
Thus, we can isolate the bottom of $\sigma\big( |A|\big)$ for $A=\wt{D}_{Q_n}$ or $A=D^{(\La)}_{Q_\infty}$ by the same circle and get
\begin{equation*}
\ket{\psi_{e;n}}\bra{\psi_{e;n}}-\dfrac{1}{\nlp{2}{\psi_e}^2}\ket{\psi_{e}}\bra{\psi_{e}}=\dfrac{1}{2i\pi}\dint_{\mathcal{C}_\eps}\dfrac{dz}{z-\wt{D}_{Q_n}}\big(\alpha R[Q_\infty-Q_n]+2\eta_n \G_n \big)\dfrac{1}{z-D^{(\La)}_{Q_\infty}}.
\end{equation*}
By dominated convergence, this operator strongly converges to $0$: this proves
\[
\nlp{2}{\psi_e}=1.
\]


\paragraph{Proof of \eqref{unif} and estimate on $E_{1,1}$}
This proof is based on the method of \cite{sok}: we know that 
\[
|m-\mu_n|\le K\alpha^2
\]
and that 
\begin{equation}\label{eq_psien}
\wt{D}_{Q_n} \psi_{e;n}=\mu_n\psi_{e;n}.
\end{equation}
In the following, we will get estimates on the Sobolev norms of $\psi_{e;n}$, this will enable us to estimate $\psh{\wt{D}_{Q_n}\psi_{e;n}}{\psi_{e;n}}$. 
We will use estimates on $g_0,g_1$ written in \eqref{estim_g}.

\subparagraph{Estimate on $\nabla \psi_{e;n}$} From \eqref{eq_psien} we have
\[
\begin{array}{l}
\nlp{2}{\D \psi_{e;n}}^2-m^2\le K\alpha^2 +4\alpha \ns{2}{Q_n}\nlp{2}{\nabla \psi_{e;n}}+4\eta_n\ns{2}{\G_n}\\
\ \ \ \ \ \ \ \ \ \ \ \ \ \ \ \ \ \ \ \ \ \ \ \ \ \ \ +2\nlp{2}{\nabla \psi_{e;n}}^2\big(\alpha \ns{2}{Q_n}^2+4\eta_n^2\ns{2}{\G_n}^2 \big)
\end{array}
\]
and $\nlp{2}{\nabla \psi_{e;n}}^2\apprle \alpha^2$.
In the same way, for $n$ sufficiently large, we can prove that
\[
\psh{\,|\nabla|^3 \psi_{e;n}}{\psi_{e;n}}\apprle \alpha^3.
\]
We multiply \eqref{eq_psien} by $|\nabla|^{1/2}$ and take the $L^2$-norm. We can drop all terms with $2\eta_n \G_n$ because all the operators that we consider are bounded in $\hl$ and $\eta_n\ns{2}{\G_n}$ tends to $0$ as $n$ tends to $+\infty$. We just have to deal with $|\nabla|^{1/2}R_{Q_n}\psi_{e;n}$. We recall that in Fourier space, the following holds \cite{ptf} 
\[
\forall\,Q\in\mathfrak{S}_2(\hl),p,q\in\RR,\ \mathscr{F}\big(R_Q;p,q\big)=\dfrac{1}{2\pi}\underset{\RR}{\dint}\dfrac{d\ell}{|\ell|^2}\wh{Q}(p-\ell,q-\ell).
\]
So, writing $\mathfrak{A}_n$ the operator whose Fourier transform is given by the integral kernel
\[
\mathscr{F}\big(\mathfrak{A}_n;p,q \big):=|p-q|^{1/2}|\wh{Q}(p,q)|,
\]
we have
\[
\Big|\mathscr{F}\big( \big[|\nabla|^{1/2},R_{Q_n}\big]\big)\Big|\le \mathscr{F}\big(R_{\mathfrak{A}_n};p,q\big).
\]
By Hardy's inequality, we have
\[
\big| \big|  \big[ |\nabla|^{1/2},R_{Q_n}\big]\psi_{e;n} \big| \big|_{L^2}\le 4\ns{2}{\,|\nabla|^{1/2}Q_n}\nlp{2}{\nabla \psi_{e;n}}\apprle \alpha^{3/2}.
\]
As 
\[
\nlp{2}{R_{Q_n}|\nabla|^{1/2} \psi_{e;n}}\le \frac{\pi}{2}\ns{2}{\,|\nabla|^{1/2}Q_n}\nlp{2}{\nabla\psi_{e;n}}\apprle \alpha^{3/2},
\]
we have $\nlp{2}{\,|\nabla|^{1/2}R_{Q_n}\psi_{e;n}}\apprle \alpha^{3/2}$ and
\begin{equation}
\psh{\,|\nabla||\D|^2\psi_{e;n}}{\psi_{e;n}}-m^\psh{|\nabla|\psi_{e;n}}{\psi_{e;n}}\apprle \alpha^3.
\end{equation}
\subparagraph{Estimates on $\chi_{e;n}$} We scale \eqref{eq_psien} by $\alpha^{-1}$, that is we consider
\[
\un{\psi_{e;n}}(x):=\alpha^{-3/2}\psi_{e;n}(\tfrac{x}{\alpha}),\ x\in\RR.
\]
This enables us to get an estimate of the lower spinor of $\psi_{e;n}$. We write
\[
\psi_{e;n}=:\begin{pmatrix} \ph_{e;n}\\ \chi_{e;n}\end{pmatrix}\in L^2(\RR,\mathbb{C}^2)^2
\]
For short we also write
\[\mathbf{g}_1(p):=g_1(p)\tfrac{p}{|p|},\ p\in\RR.
\]
We write
\[
\un{Q_n}(x,y):=\alpha^{-3}\un{Q_n}\big(\frac{x}{\alpha},\frac{y}{\alpha}\big)\text{\ and\ }\un{\G_n}(x,y):=\alpha^{-3}\un{\G_n}\big(\frac{x}{\alpha},\frac{y}{\alpha}
\]

The upper and lower spinors $\un{\ph_{e;n}}$ and $\un{\chi_{e;n}}$ of $\un{\psi_{e;n}}$ satisifies
\begin{equation}\label{scalle}
\un{\chi_{e;n}}=\dfrac{\mathbf{g}_1\big(\frac{-i\nabla}{\alpha}\big)\cdot \boldsymbol{\sigma}}{\alpha^2(\mu_n+g_0\big(\frac{)i\nabla}{\alpha}\big))}\un{\ph_{e;n}}+\big(-\alpha^2R_{\un{Q_n}}\un{\psi_{e;n}}+2\alpha \eta_n\un{\G_n}\un{\psi_{e;n}}\big)_{\downarrow}.
\end{equation}
By Hardy's inequality, we get that
\[
\nlp{2}{\chi_{e;n}}=\nlp{2}{\un{\chi_{e;n}}}\apprle \alpha.
\]
As there holds:
 \[
 \psh{-\Delta \chi_{e;n}}{\chi_{e;n}}\le \nlp{2}{\,|\nabla|^{3/2}\chi_{e;n}}\sqrt{\nlp{2}{\chi_{e;n}}\nlp{2}{\nabla \chi_{e;n}}}
\]
we also get the following (rough) estimate
\[
 \nlp{2}{\chi_{e;n}}\apprle \alpha^{4/3}.
\]

\subparagraph{Estimate on $E_{1,1}$} Using \eqref{estim_g}, we have (here $g_\star$ means $g_\star(-i\nabla)$)
\begin{align*}
\psh{\D \psi_{e;n}}{\psi_{e;n}}&=\psh{g_0\psi_{e;n}}{\psi_{e;n}}+2\mu_n\psh{\tfrac{g_1^2}{(g_0+\mu_n)^2}\phi_{e;n}}{\phi_{e;n}}+\mathcal{O}\big(\alpha(\alpha^2+\eta_n\ns{2}{\G_n})\big)\\
 &=\psh{g_0\psi_{e;n}}{\psi_{e;n}}+2m\psh{\tfrac{g_1^2}{(g_0+m)^2}\phi_{e;n}}{\phi_{e;n}}+\mathcal{O}\big(\alpha^3\big),\\
 &=m+\frac{g'_1(0)^2}{2m}\nlp{2}{\nabla \phi_{e;n}}^2+\mathcal{O}\big(\alpha^3\big).
\end{align*}
As $\psi_{v;n}=\Cha \psi_{e;n}$, we have
\[
\dfrac{1}{2}\diint \frac{|\psi_{e;n}\wedge \psi_{v;n}(x,y)|^2}{|x-y|}dxdy=D\big(|\ph_{e;n}|^2,|\ph_{e;n}|^2\big)+\mathcal{O}(\alpha^3).
\]
Using \eqref{estim_gn}, we finally get for $n$ sufficiently large
\begin{equation}
\psh{\wt{D}_{Q_n}\psi_{e;n}}{\psi_{e;n}}=m+\frac{g'_1(0)^2}{2m}\nlp{2}{\nabla \phi_{e;p}}^2-\alpha D\big(|\ph_{e;n}|^2,|\ph_{e;n}|^2\big)+\mathcal{O}(\alpha^3).
\end{equation}
As $\nlp{2}{\ph_{e;n}}^2=1-K\alpha^2$, we get
\[
 E_{1,1}\ge \mathcal{E}^0_{\text{BDF}}(Q_{\infty})=2m+\frac{\alpha^2m}{g'_1(0)^2}\mathcal{E}_{\text{CP}}\big(\wt{\ph_{e;n}}\big)+\mathcal{O}(\alpha^3),
\]
where $\mathcal{E}_{\text{CP}}$ denotes the Pekar energy \cite{LL} and $\wt{\ph_{e;n}}$ is the scaling of $\ph_{e;n}$ by $\tfrac{g'_1(0)^2}{\alpha m}$. 

We already have an upper bound of $E_{1,1}$: it has the same expansion with $\mathcal{E}_{\text{CP}}\big(\ph_{e;n}\big)$ replaced by the smallest possible value $E_{\text{CP}}$. As there holds
\[
\mathcal{E}_{\text{CP}}\big(\un{\ph_{e;n}}\big)\ge (1-\nlp{2}{\chi_{e;n}}^2)^3E_{\text{CP}}
\]
we thus have
\begin{equation}
\mathcal{E}_{\text{CP}}\big(\un{\ph_{e;n}}\big)=E_{\text{CP}}+\mathcal{O}(\alpha),
\end{equation}
and
\begin{equation}
\mu_n=m+2m\frac{\alpha^2}{g'_1(0)^2}E_{\text{CP}}+\mathcal{O}(\alpha^3).
\end{equation}

Thus $\mu_n<m$ for $\alpha$ sufficiently small. Are there other eigenvalues in $(0,m)$ ? As the Hessians are non-negative (see \eqref{hess_hess}), we have
\[
 \sigma\,|\wt{D}_{Q_n}|\subset [\mu_n-2\eta_n,+\infty)
\]
Let $\xi_n\perp \psi_n$ in $\text{Ran}\in(\pvt^n_+)$ and $s_n\in (\mu_n-2\eta_n,m)$ such that
\[
 \wt{D}_{Q_n}\xi_n=s_n\xi_n.
\]
By the same method as before used for $\psi_{e;n}$, we can prove the following:
\[
 \begin{array}{r|l}
  \nlp{2}{\nabla \xi_n}\apprle \alpha,& \nlp{2}{\,|\nabla|^{3/2}\xi_n}\apprle \alpha^{3/2},\\
  \nlp{2}{(\xi_n)_{\downarrow}}\apprle \alpha ,&\nlp{2}{\nabla (\xi_n)_{\downarrow}}\apprle \alpha^{4/3}.
 \end{array}
\]
The arrow $\downarrow$ means we take the lower spinor (which is in $L^2(\RR,\mathbb{C}^2)$). In particular we have
\begin{align*}
 s_n=\psh{\wt{D}_{Q_n}\xi_n}{\xi_n}&=m+\frac{g'_1(0)^2}{2m}\nlp{2}{\nabla \xi_n}^2-\alpha D(\xi_n^*\psi_{e;n};\xi_n^*\psi_{e;n})+\mathcal{O}(\alpha^{8/3}).
\end{align*}
\begin{remark}
We have lost $\alpha^{1/3}$ due to the rough estimate $\nlp{2}{\nabla (\xi_{n})_{\downarrow}}\apprle \alpha^{4/3}$. We can prove that this quantity is of order $\alpha^2$, but the proof is technical.
\end{remark}
\subparagraph{Estimate on $\un{\psi_{e;n}}$} We know that $\un{\psi_{e;n}}$ is close to a Pekar minimizer: its Pekar energy is
\[
 E_{CP}+\mathcal{O}(\alpha^{2/3}).
\]
For $\alpha$ sufficiently small, we know that this gives information about the distance between $\un{\psi_{e;n}}$ and the manifold $\mathscr{P}$ of Pekar minimizer \cite{lenz}:
\[
 \text{dist}_{H^1}(\un{\psi_{e;n}}, \mathscr{P})^2\le K \mathcal{E}_{\text{CP}}(\un{\psi_{e;n}})-E_{\text{CP}}.
\]
The notation $\text{dist}_{H^1}$ means the distance in the $H^1$-norm. 

This result is stated in $L^2(\RR,\mathbb{C})$, but it is not hard to prove it is also true in $L^2(\RR, \mathbb{C}^4)$: in this case $\mathscr{P}$ is isomorphic to $\RR\times \mathbb{S}^3$ (and not simply to $\RR\times \mathbb{S}^1$).

If $\un{\xi_n}$ denotes the scaling of $\xi_n$ by $\frac{g'_1(0)^2}{2\alpha m}$, there holds
\begin{equation}\label{spe_spe}
 \frac{g'_1(0)^2}{2\alpha^2m}(s_n-m)=\nlp{2}{\nabla \un{\xi_n}}^2-D\big(\un{\xi_n}^*\un{\psi_{e;n}},\un{\xi_n}^*\un{\psi_{e;n}}\big)+\mathcal{O}(\alpha^{2/3}).
\end{equation}
Eventually by replacing $\un{\psi_{e;n}}$ by its projection $\phi_{\text{CP}}^n$ onto $\mathscr{P}$, we also have
\begin{equation}
 \frac{g'_1(0)^2}{2\alpha^2m}(s_n-m)=\nlp{2}{\nabla \un{\xi_n}}^2-D\big(\un{\xi_n}^*\phi_{\text{CP}}^n,\un{\xi_n}^*\phi_{\text{CP}}^n\big)+\mathcal{O}(\alpha^{1/3}).
\end{equation}
\subparagraph{Proof of \eqref{unif}} We just have to study the spectrum of $\sigma(-\Delta -R\big( \ket{\phi_{\text{CP}}^n}\bra{\phi_{\text{CP}}^n}\big))$, and precisely its negative eigenvalues. Its smallest eigenvalue is $E_{\text{CP}}$ with eigenvector $\phi_{\text{CP}}^n$. Now we seek the second smallest eigenvalue, that is
\begin{equation}
 F_{\text{CP}}:=\inf\Big\{ \psh{\big(-\Delta -R\big( \ket{\phi_{\text{CP}}^n}\bra{\phi_{\text{CP}}^n}\big)\big)f}{f},\ f\perp \phi_{\text{CP}}^n\in H^1,\ \nlp{2}{f}=1 \Big\}.
\end{equation}
By studying a minimizing sequence, we get
\begin{equation}
 F_{\text{CP}}>E_{\text{CP}}.
\end{equation}
By continuity the same holds for the spectrum of $-\Delta -R\big(\ket{\un{\psi_{e;n}}} \bra{\un{\psi_{e;n}}}\big)$: for $\alpha$ sufficiently small (and $n$ sufficiently big) its smallest eigenvalue $t_n$ has multiplicity one and its second smallest eigenvalue $\wt{t}_n$ is away from $t_n$, uniformly in $\alpha$ (and $n$):
\[
 \wt{t}_n-t_n>\frac{F_{\text{CP}}-E_{\text{CP}}}{2}>0.
\]

As a consequence, we get from \eqref{spe_spe} the following:
\begin{equation}
 s_n-\mu_n\ge \frac{\alpha^2 m}{g'_1(0)^2}\big(F_{\text{CP}}-E_{\text{CP}}\big)+\mathcal{O}(\alpha^{7/3}),
\end{equation}
and \eqref{unif} holds.

\subsection{Proof of Theorems \ref{foform} and \ref{non_rel}}

In fact, it suffices to follow the proof of Theorem \ref{critic}: instead of having an almost minimizer, we deal with a real minimizer $\ov{P}=\ov{Q}+\PP$. Technically speaking, we just have to drop the term $\eta_n\G_n$ in the equations and by the same method we prove the following.
\begin{enumerate}
	\item There exist $0<\mu<m$ and a wave function $\psi_e\in \hl$ such that
	 \begin{equation}
	 	\left\{\begin{array}{l}
			\ov{P}=\ket{\psi_e}\bra{\psi_e}-\ket{\Cha \psi_e}\bra{\Cha \psi_e}+\chi_{(-\infty,0)}\big( \Pi_\La D_{\ov{Q}} \Pi_\La\big),\\
			\Pi_\La D_{\ov{Q}} \Pi_\La \psi_e=\mu\psi_e.
		\end{array}\right.
	 \end{equation}
	 \item We have $\nlp{2}{\,|\nabla|^3 \psi_e}\apprle \alpha^{3/2}$. Splitting $\psi_e$ into upper and lower spinors $\ph_e$ and $\chi_e$, we have $\nlp{2}{\chi_e}\apprle \alpha$. We write $\wt{\ph_e}(x):=\la^{3/2}\ph_e(\la x)$ with $\la=\tfrac{g'_1(0)^2}{\alpha m}$. The following holds:
	 \begin{equation}
	 	\left\{
		\begin{array}{rl}
			E_{1,1}&=2m+\frac{\alpha^2m}{g'_1(0)^2}\mathcal{E}_{\text{CP}}\big(\wt{\ph_e}\big)+\mathcal{O}(\alpha^3)\\
				    &=2m+\frac{\alpha^2m}{g'_1(0)^2}\mathrm{E}_{\text{CP}}+\mathcal{O}(\alpha^3),\\
			\mu&=m+2m\frac{\alpha^2}{g'_1(0)^2}E_{\text{CP}}+\mathcal{O}(\alpha^3).
		\end{array}
		\right.
	 \end{equation}
	 \item In the limit $\alpha\to 0$ we have
	 \[
	 \lim_{\alpha\to 0}\nlp{2}{\chi_e}=0\text{\ and\ }\lim_{\alpha\to 0}\mathcal{E}_{\text{CP}}\big(\wt{\ph_e}\big)=\mathrm{E}_{\text{CP}}.
	 \]
\end{enumerate}

The \emph{geometrical} description of a minimizer of Theorem \ref{foform} has already been proved at the end of Subsection \ref{subscritic} under the assumption of existence.

\section{Proofs on results on the variational set}\label{proofmanif}

\subsection{On the manifold $\mathscr{M}$: Theorem \ref{structure}, Propositions \ref{manim}, \ref{gragra}}

\noindent\textbf{Proof of Theorem \ref{structure}}

\noindent -- As $Q$ is a compact self-adjoint operator, we apply the spectral theorem and write
\[
Q=\ssum_{i\in \mathbb{Z}^*}\mu_i \ket{b_i}\bra{b_i},
\]
where $(\mu_i)_{i\in\mathbb{N}}$ (resp. $(\mu_i)_{i\in\mathbb{Z}_-^*}$) is the non-increasing sequence of positive eigenvalues of $Q$ (resp. increasing sequence of negative eigenvalues).

It is clear that $-1\le Q\le 1$. If $Q\psi=\psi$, then necessarily $P_1\psi=\psi$ and $P_0\psi=0$, analogously if $Q\psi=-\psi$, then $P_1\psi=0$ and $P_0 \psi=\psi$. 

Up to index translation  we have:
\begin{equation}\label{forme}
A:=Q-\Big\{\ssum_{i=1}^{M_+}\ket{a_i}\bra{a_i}-\ssum_{i=1}^{M_-}\ket{a_{-i}}\bra{a_{-i}}\Big\}=\ssum_{i\in \mathbb{Z}^*}\mu_i \ket{b_i}\bra{b_i}=A_p-A_n,
\end{equation}
where $A_p$ is the sum over positive $i$ and $-A_n$ over negative $i$. 

\begin{notation}
For short, for any $\mu\in \mathbb{R}$ and any self-adjoint operator $S$, we write $E^S_{\mu}=\text{Ker}(S-\mu)$ the spectral subspace of $S$.  

Furthermore, for an operator $B$ we write
\[
 B^{\eps_1\ \eps_2}=P_0(\eps_1)B P_0(\eps_2),\ \eps_i=\pm,\ P_0(-)=P_0,\ P_0(+)=1-P_0.
\]
\end{notation}
\noindent -- We know that
\[
Q^{++}-Q^{--}=Q^2=\ssum_{i=1}^{M_+}\ket{a_i}\bra{a_i}+\ssum_{i=1}^{M_-}\ket{a_{-i}}\bra{a_{-i}}+\ssum_{i\in \mathbb{Z}^*}\mu_i^2 \ket{b_i}\bra{b_i}.
\]
In particular $[Q^2,P_0]=0$ and all the spectral subspaces of $Q^2$ are $P_0$-invariant. For any $\mu>0$, 
\[
E_{\mu^2}^{Q^2}=E^Q_{\mu}\overset{\perp}{\oplus} E^Q_{-\mu}=E^{Q^{++}}_{\mu^2}\overset{\perp}{\oplus} E^{Q^{--}}_{-\mu^2}.
\]
For $i\in\mathbb{N}$, let $c_i$ be a unitary eigenvector for $Q^{++}$ with eigenvalue $0<\mu_i^2<1$. We write
\[
c_i=c_p+c_n,\ c_p\in\text{Ran}(A_p),\ c_n\in\text{Ran}(A_n).
\]
We have $A_p c_p=\mu_i c_p$ and $A_n c_n=-\mu_i c_n$. Moreover $c_n\neq 0$, otherwise $(1-P_0) c_p=c_p$ and
\[
A c_p=\mu_i c_p=((1-P_0) -(1-P_1))c_p\ i.e.\ (1-P_1)c_p=(1-\mu_i)c_p.
\]
This would give $\mu_i=1$ or $\mu_i=0$. By the same argument $c_p\neq 0$. We have $P_0 c_p=-P_0 c_n$ and this vector is non-zero, otherwise $(1-P_0) c_p=c_p$. Thus the two-dimensional plane $\Pi=\text{Span}(c_p,c_n)$ is in $E^{Q^2}_{\mu_i^2}$ and there exists an orthonormal basis $(e_+=c_i,e_-)$ of $\Pi$ such that $P_0 e_-=e_-$ (and $(1-P_0) c_i=c_i$).

We write $c_p=||c_p|| d_p$ and $c_n=||c_n||d_n$ and up to a phase, we have:
\[
c_i=\cos(\phi)d_p+\sin(\phi)d_n.
\]
There holds:
\[
Q^2 c_i=\mu_i^2c_i=A_p c_i=\mu_i(1-P_0)(\cos(\phi)d_p+\sin(\phi)d_n)=\mu_i(\cos(\phi)^2-\sin(\phi)^2)c_i,
\]
and $\mu_i=\cos(2\phi)$. We have 
\[
E^{Q^2}_{\mu_i^2}=\Pi\overset{\perp}{\oplus}R.
\]
\noindent -- By induction over the dimension of the remainder $\text{Dim}(R\cap E^{Q^{++}}_{\mu_i^2})$, we can decompose $E^{Q^2}_{\mu_i^2}$ as a sum of orthogonal planes: by symmetry there holds $\text{Dim}\,E^{Q^{++}}_{\mu_i^2}=\text{Dim}\,E^{Q^{--}}_{\mu_i^2}$.
Each plane $\Pi$ is invariant under the action of $Q$ and $\PP$ and so also under that $P=Q+\PP$. Therefore, there also exists an orthonormal basis $(v_+,v_-)$ of $\Pi$ such that $P_1v_-=v_-$ and $(1-P_1)v_+=v_+$. Up to a phase we suppose that 
\begin{equation}
v_-=\cos(\theta)e_-+\sin(\theta)e_+\text{\ and\ }v_+=-\sin(\theta)e_-+\cos(\theta)e_+,\ \theta\in(0,\tfrac{\pi}{2}).
\end{equation}
In the plane $\Pi$ we thus have:
\[
Q|_{\Pi}=\ket{v_-}\bra{v_-}-\ket{e_-}\bra{e_-}.
\]
Such an operator has eigenvalues $\pm\sin(\theta)$ with eigenvectors
\begin{equation}\label{eigg}
\left\{\begin{array}{ll}
f_+=\sqrt{\frac{1-\sin(\theta)}{2}}e_-+\sqrt{\frac{1+\sin(\theta)}{2}}e_+&\text{associated\ to\ }\sin(\theta),\\
f_-=-\sqrt{\frac{1+\sin(\theta)}{2}}e_-+\sqrt{\frac{1-\sin(\theta)}{2}}e_+&\text{associated\ to\ }-\sin(\theta)
\end{array}
\right.
\end{equation}
\hfill{\small$\Box$}

\noindent \textbf{Proof of Proposition \ref{manim}}
In general, let $P_1$ and $P_2$ be two orthogonal projectors in $\hl$. If $P_2=UP_1U^{-1}$ where $U$ is a unitary operator, we have:
\begin{equation}\label{Uiff}
 P_2-P_1\in\mathfrak{S}_2(\hl)\ \iff\ [U,P_1]U^{-1}\in\mathfrak{S}_2(\hl)\ i.e.\ [U,P_1]\in\mathfrak{S}_2(\hl).
\end{equation}

\noindent -- For any $P_1\in\mathscr{M}$ and any $P_2\in\mathscr{M}$ with $\nb{P_1-P_2}<1$, we can decompose $P_2-P_1$ as in Theorem \ref{structure} but with $P_1$ as new reference (the decomposition is the same but with $e_j\in\text{Ran}\,(1-P_1)$ and $e_{-j}\in\text{Ran}\,P_1$):
\[
\left\{\begin{array}{l}
P_2-P_1=\ssum_{j\in\mathbb{N}}(\ket{v_{-j}}\bra{v_{-j}}-\ket{e_{-j}}\bra{e_{-j}}),\ v_{-j}=\cos(\theta_j)e_{-j}+\sin(\theta_j)e_j\\
P_2 v_{-j}=v_{-j},\ P_1e_{-j}=e_{-j},P_1e_j=0\text{\ and\ }\ssum_{j\in\mathbb{N}}\sin(\theta_j)^2<+\infty.
\end{array}\right.
\]
Above we have $\theta_j\in(0,\tfrac{\pi}{2})$ for all $j\in\mathbb{N}$.
Let $A$ be defined as follows:
\[
A=\ssum_{j\in\mathbb{N}}\theta_j (\ket{e_j}\bra{e_{-j}}-\ket{e_{-j}}\bra{e_{j}}),\ \theta_j\in (0,\tfrac{\pi}{2}),
\]
then we have $P_2=e^AP_1e^{-A}$, $A^*=-A$ and
\begin{equation}
[A,P_1]=\ssum_{j\in\mathbb{N}}\theta_j (\ket{e_j}\bra{e_{-j}}+\ket{e_{-j}}\bra{e_{j}})\in\mathfrak{S}_2(\hl).
\end{equation}
Furthermore $[\exp(A),P_1]\in\mathfrak{S}_2(\hl)$: for all $k\in\mathbb{N}$, there holds:
\[
[A^k,P_1]=\ssum_{j=0}^{k-1}A^j[A,P_1]A^{k-1-j},
\]
and
\begin{equation}\label{exp(A)}
\ns{2}{[\text{exp}(A),P_1]}\le \ssum_{k=1}^{+\infty}\frac{1}{k!}\big\{k\ns{2}{[A,P_1]}\nb{A}^{k-1}\big\}=\ns{2}{[A,P_1]}\text{exp}\,\nb{A}.
\end{equation}
Let us call this $A$ the \emph{canonical} antiunitary operator $L_{P_1}(P_2)$ associated to $P_2$: we will see it does not depend on the choice of eigenvectors $e_j$.
\begin{remark}\label{formexp}
 In the case $\nb{P_2-P_1}=1$, we have $1,-1\in\sigma(P_2-P_1)$: indeed $P_2-P_1$ may be decomposed as in \eqref{forme} with $M_+=M_-$ because $\text{Tr}(P_2-P_1)=0$. 
 
 We still have $P_2=e^{A}P_1e^{-A}$ with
 \begin{equation}\label{eqformexp}
 A=\ssum_{i=1}^{M_+}\frac{\pi}{2}\Big(\ket{a_i}\bra{a_{-i}}-\ket{a_{-i}}\bra{a_{i}}\Big) +\ssum_{j\ge 1}\theta_j \Big(\ket{e_j}\bra{e_{-j}}-\ket{e_{-j}}\bra{e_{j}}\Big),
 \end{equation}
where $a_i,e_j\in\text{Ran}(1-P_1)$ and $a_{-i},e_{-j}\in\text{Ran}\,P_1$ form an orthonormal family as in the decomposition of Theorem \ref{structure} (in particular the non-zero eigenvalues in $(-1,1)$ are the $\pm\sin(\theta_i)$).
\end{remark}


\noindent -- Let $(\mathfrak{m}_{P_1},\ns{2}{\cdot})$ be the set of compact operators:
\[
\mathfrak{m}_{P_1}:=\{a\in\mathcal{B}(\hl),\ ((1-P_1)a P_1)^*=-P_1a (1-P_1)\in\mathfrak{S}_2(\hl),\ (1-P_1)a (1-P_1)=P_1 P_1=0\}.
\]
\begin{remark}
As we consider operators in $\mathcal{B}(\hl)$ we can replace $1$ by $\Pi_\La$ in the definition.
\end{remark}

The map $\Phi_{P_1}$ 
\begin{equation}\label{PhiP}
 \Phi_{P_1}:\begin{array}{rll}
          (\mathfrak{m}_{P_1},0)&\longrightarrow& (\mathscr{M}, P_1)\\
          a&\mapsto & e^a P_1 e^{-a}
         \end{array}
\end{equation}
is differentiable and we have:
\[
\forall A\in\mathfrak{m}_{P_1},\ \dd \Phi_{P_1}(P_1)\cdot A=[A,P_1].
\]
This map
\[
\dd \Phi_{P_1}:\mathfrak{m}_{P_1}\to \{[A,P_1],\ A\in\mathfrak{m}_{P_1}\}=:\text{Ran}( \dd \Phi_{P_1})
\]
is invertible with inverse
\[
\dd \Phi_{P_1}^{-1}:v\in\text{Ran}( \dd \Phi_{P_1})\mapsto [v,P_1]\in\mathfrak{m}_{P_1}.
\]
This proves that in a neighbourhood of $P_1$, the corresponding part of $\mathscr{M}$ is the graph of some function $\mathcal{F}_{P_1}$.

Indeed, if we see the set
\[
\PP+\mathfrak{S}_2(\hl)=P_1+\mathfrak{S}_2(\hl)
\]
as an affine space with associated vector space $\mathfrak{S}_2(\hl)$, then we have
\[
\mathfrak{S}_2(\hl)=\mathfrak{m}_{P_1}\overset{\perp}{\oplus}\text{Ran}(\dd \Phi_{P_1})\overset{\perp}{\oplus}\{u\in\mathfrak{S}_2(\hl),\ P_1 u(1-P_1)=(1-P_1)uP_1=0\}.
\]
We decompose any $Q\in\mathfrak{S}_2(\hl)$ with respect to $\text{Ran}(\dd \Phi_{P_1})\oplus (\text{Ran}(\dd \Phi_{P_1}))^{\perp}$:
\[
Q=v[P_1;Q]+w[P_1;Q]\in \text{Ran}(\dd \Phi_{P_1})\oplus (\text{Ran}(\dd \Phi_{P_1}))^{\perp}.
\]
In a neighbourhood $\mathcal{V}_{P_1}$ of $P_1$, the set  $\mathcal{V}_{P_1}\cap \mathscr{M}$ is a portion of the graph of
\[
\mathcal{F}_{P_1}:v\in \text{Ran}( \dd \Phi_{P_1})\mapsto P_1+w\big[P_1; e^{[v,P_1]} P_1 e^{-[v,P_1]} -P_1\big]\in P_1+(\text{Ran}\, \dd \Phi_{P_1})^{\perp}.
\]
\noindent -- Thus for any $P_1\in\mathscr{M}$, there exists a neighbourhood $\mathcal{V}_{P_1}\ni P_1$ such that $\mathscr{M}\cap\mathcal{V}_{P_1}$ is a manifold with $\text{T}_{P_1}\mathscr{M}=\text{Ran}( \dd \Phi_{P_1})$.  To conclude $\mathscr{M}$ is a proper manifold, it suffices to compare the neighbourhood of $\mathscr{M}$ (or prove that $\mathscr{M}$ is connected): for $P_1,P_3\in\mathscr{M}$, we use Remark \ref{formexp} and write $P_3=e^A P_1 e^{-A}$ with $A\in\mathfrak{m}_{P_1}.$ Then it is clear that the map
\[
\mathfrak{T}(P_1,P_3):\begin{array}{rll}
 (\mathscr{M},P_1)&\longrightarrow& (\mathscr{M},P_3)\\
 P&\mapsto& e^A P e^{-A}
 \end{array}
\]
is an isometry and that its differential $\mathfrak{t}(P_1,P_3)$ is an isometry that maps $\text{T}_{P_1}\mathscr{M}$ onto $\text{T}_{P_3}\mathscr{M}$.  The map $t\in [0,1]\mapsto e^{tA}P_1 e^{-tA}\in\mathscr{M}$ links $P_1$ and $P_3$.

Moreover the map
\[
L_{P_1}:\begin{array}{cll} \{P\in \mathscr{M},\ \nb{P-P_1}<1\}&\longrightarrow& \mathfrak{m}_{P_1}\\
P&\mapsto& A
\end{array}
\]
is locally invertible around $P_1$ with (local) inverse $\Phi_{P_1}$.

More generally, we can prove that the restriction of $\Phi_{P_1}$ to the $a\in\mathfrak{m}_{P_1}$ with $\nb{a}<\tfrac{\pi}{2}$ is one-to-one: it suffices to consider the spectral decomposition of $a$ and link spectral subspaces with rotations.



\hfill{\small$\Box$}

\noindent\textbf{Proof of proposition \ref{gragra}}
\begin{remark}
\begin{enumerate}
 \item We recall that if $P_1$ and $P_2$ are two projectors such that $P_1-P_2$ is Hilbert-Schmidt, then
 \begin{equation}\label{ptrace}
  A\in\mathfrak{S}_1^{P_1}\iff A\in\mathfrak{S}_1^{P_2}\text{\ and\ }\text{Tr}_{P_1}(A)=\text{Tr}_{P_2}(A).
 \end{equation}
 \item For any $A\in\mathcal{B}$ and any projector $P$ we have:
 \begin{equation}\label{ccom}
  [[A,P],P]=(1-P)AP+PA(1-P).
 \end{equation}
\end{enumerate}
\end{remark}

If we restrict $\mathcal{E}_{\text{BDF}}$ to $\mathscr{M}$, using \eqref{ptrace} and \eqref{ccom} we get that for $(P,v)\in\text{T}\mathscr{M}$:
\begin{equation}\label{difftaninc}
\text{d}\mathcal{E}_{\text{BDF}}^0(P)\cdot v=\text{Tr}_P(\Pi_\Lambda D_{P-\PP} \Pi_\Lambda v)=\text{Tr}_P\big([[\Pi_\Lambda D_{P-\PP}\Pi_\Lambda ,P],P]v\big).
\end{equation}
We write $Q=P-\PP$, $\boldsymbol{\pi}=\chi_{(-\infty,0)}(\Pi_\Lambda D_Q \Pi_\Lambda)$ and $\G =P-\boldsymbol{\pi}$. We have:
\begin{align*}
 P \Pi_\Lambda D_Q\Pi_\La (1-P)&=(\boldsymbol{\pi}+\G )\Pi_\La D_Q \Pi_\La (1-\boldsymbol{\pi}-\G),\\
                               &=\boldsymbol{\pi} -\Pi_\La D_Q \Pi_\La \G +\G \Pi_\La D_Q\Pi_\La (1-\boldsymbol{\pi})- \G \Pi_\La D_Q\Pi_\La \G.
\end{align*}
Thus
\begin{equation}
 [[\Pi_\La D_Q\Pi_\La, P],P]= |\Pi_\La D_Q \Pi_\La|\G+\G |\Pi_\La D_Q \Pi_\La|-2\G \Pi_\La D_Q \Pi_\La \G.
\end{equation}
We have:
\begin{align*}
 |\Pi_\La D_Q\Pi_\La|^2&= \Pi_\La (\D)^2+\alpha \big(\Pi_\La B_Q\Pi_\La\D+\D\Pi_\La B_Q\Pi_\La\big)+\alpha^2(\Pi_\La B_Q\Pi_\La)^2\\
                       &\le \Pi_\La (\D)^2\Big(1+\alpha \nb{\Pi_\La B_Q\Pi_\La \text{inv}(\D)}\Big)^2,\\
                       &\le \Pi_\La (\D)^2\Big(1+\alpha K \nb{V_Q \tfrac{\Pi_\La}{\sqrt{1-\Delta}}}+\nb{R_Q \tfrac{\Pi_\La}{\sqrt{1-\Delta}}}\Big)^2.
\end{align*}
We have $\G=(P-\PP)+(\PP-\boldsymbol{\pi})\in\mathfrak{S}_1^{\PP}(\hl)$. So the following holds:
\[
\big| \big|\,|\Pi_\La D_Q \Pi_\La|\G\big| \big|_{\mathfrak{S}_2}\apprle E(\La)^{1/2}\ns{2}{\,|\D|^{1/2}\G}(1+\alpha (\sqrt{D(\rho_Q,\rho_Q)}+\ns{2}{|\D|^{1/2} Q}))^2,
\]
and
\[
 \ns{2}{\G \Pi_\La D_Q \Pi_\La \G}\le 2 \big| \big|\,|\Pi_\La D_Q\Pi_\La|^{1/2}\G\big| \big|_{\mathfrak{S}_2}^2<+\infty.
\]
\hfill{\small$\Box$}

\subsection{On the manifold $\mathscr{M}_{\mathscr{C}}$: Propositions \ref{manicsym}, \ref{conn} and \ref{chasym}}

\noindent \textbf{Proof of Proposition \ref{manicsym}}

\noindent Let $P_1,P_2\in\mathscr{M}_{\mathscr{C}}$ such that $\nb{P_2-P_1}<1$. Thanks to Theorem \ref{structure}, we know that $P_2$ can be written as $P_2=e^A P_1 e^{-A}$ where $A\in\mathcal{B}(\hl)$ is antiunitary and 
\[
P_1 A P_1=(1-P_1)A (1-P_1).
\]

\noindent -- Taking into account the $\Cha$-symmetry we can say more: thanks to \eqref{chaisom} we can follow the proof of Proposition \ref{chasym} with $\PP$ replaced by $P_1$. This gives
\begin{equation}\label{chaa}
 \Cha A \Cha=A.
\end{equation}
Indeed there exist $\mathcal{J}\subset \mathbb{Z}^*$ with $-\mathcal{J}=\mathcal{J}$ and $(e_j)_{j\in J}$ in $\hl^{\mathcal{J}}$ such that 
\begin{enumerate}
\item $(e_j)_j\cup (Ce_j)_j$ is an orthonormal basis for $\text{Ran}(P_2-P_1)$,
\item for all $j\in \mathcal{J}$, $j>0$: $P_1e_j=0$ and $P_1 e_{-j}=e_{-j}$,
\item each $4$-dimensional space $\text{Span}(e_j,e_{-j},\Cha e_j,\Cha e_{-j})$ is spanned by four eigenvectors $f_j\perp\Cha f_{-j}$ with eigenvalue $\sin(\theta_i)>0$ and $f_{-j}\perp\Cha f_j$ with eigenvalue $-\sin(\theta_i)$.
\end{enumerate}
Then $A$ is defined as follows:
\[
 A=\ssum_{j\in \mathcal{J}}\theta_j \Big(\ket{e_j}\bra{e_{-j}}-\ket{e_{-j}}\bra{e_j}-\ket{\Cha e_{-j}}\bra{\Cha e_j}+\ket{\Cha e_{j}}\bra{\Cha e_{-j}}\Big) 
\]
It is easy to check \eqref{chaa} from this formula. Reciprocally, let  $A\in\mathfrak{m}_P$ be an antiunitary map satisfying \eqref{chaa}. Then we know that $e^{A}Pe^{-A}\in\mathscr{M}$. Moreover we have $-\Cha e^{A}\Cha =-e^A.$ It follows that
\begin{align*}
-\Cha (e^{A} P e^{-A}-P) \Cha&=\Cha e^{A} \Cha (-\Cha P \Cha) \Cha e^{-A}\Cha+\Cha P \Cha,\\
        &=e^A(-(\Pi_\La-P) )e^{-A} +(\Pi_\La-P),\\
        &=-\Pi_\La+e^A P e^{-A}+\Pi_\La-P=e^A P e^{-A}-P.
\end{align*}
In other words $e^A P e^{-A}\in\mathscr{M}_{\mathscr{C}}$.
 Thus $\Phi_{P_1}$ (\emph{cf} \eqref{PhiP}) is a local isomorphism from $(\mathfrak{m}_{P_1},0)$ to $(\mathscr{M},P_1)$, and its restriction
 \[
  \Phi_{P_1}^{\mathscr{C}}:\begin{array}{rll}\mathfrak{m}_{P_1}^{\mathscr{C}}&\longrightarrow&\mathscr{M}_{\mathscr{C}}\\
                              a&\mapsto& e^a P_1 e^{-a}
               \end{array}
 \]
 is well-defined and is a local isomorphism from $(\mathfrak{m}_{P_1}^{\mathscr{C}}, 0)$ to $(\mathscr{M}_{\mathscr{C}},P)$.
There remains to prove that for any $P_1,P_2\in \mathscr{M}_{\mathscr{C}}$, there exists an isometry of $\mathfrak{S}_2$, that maps $\mathfrak{m}_{P_1}^{\mathscr{C}}$ onto $\mathfrak{m}_{P_2}^{\mathscr{C}}$. If $\nb{P_1-P_2}<1$, this isometry is given by
\[
 \phi_{\mathscr{C}}^0(P_1,P_2): X\in\mathfrak{S}_2(\hl)\mapsto \text{exp}(L_{P_1}(P_2))\,X\,\text{exp}(-L_{P_1}(P_2))\in\mathfrak{S}_2(\hl).
\]
The restriction is:
\[
 \phi_{\mathscr{C}}(P_1,P_2):X\in \mathfrak{m}_{P_1}^{\mathscr{C}}\mapsto \text{exp}(L_{P_1}(P_2))\,a\,\text{exp}(-L_{P_1}(P_2)),
\]
indeed, as $\Cha L_{P_1}(P_2)\Cha= L_{P_1}(P_2)$ we have $\Cha \phi_{\mathscr{C}}(P_1,P_2;a)\Cha= \phi_{\mathscr{C}}(P_1,P_2;a)$. If $\nb{P_1-P_2}=1$ then we can write
\[
 P_2-P_1=\ssum_{k=1}^K\big(\ket{a_k}\bra{a_k}-\ket{\Cha a_k}\bra{\Cha a_k}\big)+\g(P_1,P_2),
\]
where $(a_k)_k\cup (\Cha a_k)_k$ is an orthonormal family which is orthogonal to $\text{Ran}\,\g(P_1,P_2)$ and $\nb{\g(P_1,P_2)}<1$. We also have $P_1 \Cha a_k=\Cha a_k$ and $P_1 a_k=0$. We define
\[
\left\{
\begin{array}{rll}
 P_{12}&:=&P_1+\ssum_{k=1}^K\big(\ket{a_k}\bra{a_k}-\ket{\Cha a_k}\bra{\Cha a_k}\big)\in\mathscr{M}_{\mathscr{C}},\\
 U_{12}&:=&\ssum_{k=1}^K\big(\ket{\Cha a_k}\ket{a_k}-\ket{a_k}\bra{\Cha a_k}\big)\in\mathbf{U}(\hl).
 \end{array}\right.
\]
Then $\nb{P_2-P_{12}}<1$ and $U_{12}P_1 U_{12}^*=-U_{12}P_1 U_{12}=P_{12}$. Moreover
\[
 \phi_{\mathscr{C},P_1,P_{12}}:\begin{array}{rll}
                                \mathfrak{m}_{P_1}^{\mathscr{C}}&\longrightarrow& \mathfrak{m}_{P_{12}}^{\mathscr{C}}\\
                                a&\mapsto& U_{12} a U_{12}^{-1}
                               \end{array}
\]
is well-defined and is an isometry. Indeed, as $\Cha U_{12}\Cha=-U_{12}$, we get that 
\[
\Cha U_{12} a U_{12}^{-1} \Cha=U_{12} a U_{12}^{-1}.
\]
This proves the isometric isomorphisms
\[
\begin{array}{|rclcl}
\mathfrak{S}_{2}(\hl)&\underset{\phi_{\mathscr{C}}^0(P_1,P_{12})}{\overset{\simeq}{\longrightarrow}}&\mathfrak{S}_{2}(\hl)&\underset{\phi_{\mathscr{C}}^0(P_{12},P_{2})}{\overset{\simeq}{\longrightarrow}}&\mathfrak{S}_{2}(\hl),\\
\mathfrak{m}_{P_1}^{\mathscr{C}}&\underset{\phi_{\mathscr{C}}(P_1,P_{12})}{\overset{\simeq}{\longrightarrow}}&\mathfrak{m}_{P_{12}}^{\mathscr{C}}&\underset{\phi_{\mathscr{C}}(P_{12},P_{2})}{\overset{\simeq}{\longrightarrow}}&\mathfrak{m}_{P_{2}}^{\mathscr{C}}.
\end{array}
\]

So $\mathscr{M}_{\mathscr{C}}$ is a submanifold and the characterization of the tangent planes \eqref{tangentc} follows from that of $\mathscr{M}$.

\noindent -- Let us show that $\mathscr{M}_{\mathscr{C}}$ is invariant under the flow of $\mathcal{E}_{\text{BDF}}^0$: it suffices to show that for any $P\in\mathscr{M}_{\mathscr{C}}$, the gradient $\nabla \mathcal{E}_{\text{BDF}}^0(P)$ (\emph{cf} \eqref{defgradient}) is in $\text{T}_P \mathscr{M}_{\mathscr{C}}$. For a $\Cha$-symmetric state $P$, we write $Q:=P-\PP$.

That the density $\rho_{Q}$ vanishes is clear from \eqref{calculcha} and the fact that for any $\psi\in \hl$ and $x\in\mathbb{R}^3$ we have $|\Cha \psi(x)|^2=|\psi(x)|^2$. From \eqref{chargeconjprec}, we get that for $-\Cha Q\Cha=Q$ there holds:
\[
 -\Cha Q\Cha (x,y)=Q(x,y)\text{\ so\ }-\Cha R_Q \Cha(x,y)=R_Q(x,y)=\frac{Q(x,y)}{|x-y|}.
\]
As $-\Cha \D \Cha=\D$, it follows that:
\begin{equation}
 -\Cha \big(\D+\alpha (\rho_Q*\tfrac{1}{|\cdot|}-R_Q)\big)\Cha= -\Cha (\D-\alpha R_Q)\Cha=\D-\alpha R_Q.
\end{equation}
We remark that $[\Pi_\La,\Cha]=0$, and $\Cha P \Cha=1-P$ and $\Cha (1-P)\Cha =P$. Thus

\begin{align*}
 -\Cha \big[ \big[\Pi_\La D_Q \Pi_\La; P\big];P \big]\Cha&=-\Cha \big( P\Pi_\La D_Q \Pi_\La(1-P)+ (1-P)\Pi_\La D_Q \Pi_\La P\big)\Cha\\
                                                          &=(1-P)\big(-\Pi_\La \Cha D_Q \Cha \Pi_\La\big)P+P\big(-\Pi_\La \Cha D_Q \Cha \Pi_\La\big)(1-P)\\
                                                          &=(1-P)\Pi_\La D_Q \Pi_\La P+P\Pi_\La D_Q \Pi_\La (1-P)\\
                                                          &=\big[ \big[\Pi_\La D_Q \Pi_\La; P\big];P \big].
\end{align*}


\medskip

\noindent \textbf{Proof of Proposition \ref{conn}}
 Let $c:t\in[0,1]\mapsto c(t)\in \mathscr{M}_{\mathscr{C}}$ be a continuous map such that $c(0)=0$ and $\nb{c(1)}=1$. By Theorem \ref{structure} and Proposition \ref{chasym}, any $c(t)$ has the following form:
  \[
  \begin{array}{l}
   c(t)=\ssum_{j\in\mathbb{N}}\lambda_j(\ket{f_{j}(t)}\bra{f_{j}(t)}-\ket{f_{-j}(t)}\bra{f_{-j}(t)}+\ket{\Cha f_{-j}(t)}\bra{\Cha f_{-j}(t)}-\ket{\Cha f_j(t)}\bra{f_j(t)})\\
   \ \ \ +\ssum_{j=1}^{N(t)}(\ket{a_j(t)}\bra{a_j(t)}-\ket{\Cha a_j(t)}\bra{\Cha a_j(t)}),
  \end{array}
  \]
where $(a_j)_j\cup(\Cha a_j)_j\cup(f_j)_j\cup(\Cha f_j)$ is an orthonormal family and $(\la_j)_j$ is the sequence of positive eigenvalues lesser than $1$. Each plane $\text{Span}(f_j,f_{-j})$ (resp. $\text{Span}(\Cha f_j,\Cha f_{-j})$) is spanned by $e_j\in\text{Ran}(\PPP)$ and $e_{-j}\in\text{Ran}(\PP)$ (resp. $\Cha e_{-j}\in \text{Ran}(\PPP)$ and $\Cha e_{j}\in \text{Ran}(\PP)$).

Let $t_0$ be $\inf\{t\in[0,1],\ \nb{c(t)}=1\}.$ For any $t\in[0,1]$ and any $\mu\in\sigma(c(t))\backslash\{1,0\}$, $4\,|\ \text{Dim}\,E^{c(t)^2}_{\mu^2}$. In particular, for $t<t_0$ the number
\[
 J(c(t))=\text{Dim}\,\underset{\tfrac{1}{2}<\mu\le 1}{\bigoplus}E^{c(t)^2}_{\mu^2}\text{\ is\ divisible\ by\ }4.
\]
By continuity, $J(c(t))$ is divisible by $4$ for any $t$ : the variations of $J$ follow the variations of the $\la_i'$s ($\la_i$ equals $\sin(\widehat{\mathbb{C} v_j,\,\mathbb{C}e_{j}})$ in the notations of Theorem \ref{structure}). Such an eigenvalue is associated to $4$-dimensional spaces of type $\text{Span}(f_j,f_{-j},\Cha f_j,\Cha f_{-j})$ and each of them has a basis made of four eigenvectors in $E^{c(t)^2}_{\la_i^2}$.

Thus $4\,|\ J(c(1))$ and for any unitary $\psi\in\text{Ran}\,\PPP$, there is no continuous path in $\mathscr{M}_{\mathscr{C}}$ that links $0$ and $Q_{\psi}=\ket{\psi}\bra{\psi}-\ket{\Cha\psi}\bra{\Cha\psi}$.
It is then straightforward to prove that for any $\g\in\mathscr{M}_{\mathscr{C}}$, if $4\,|\ J(\g)$ then there exists a path that links $0$ and $\g$ else there exists a path that links $Q_{\psi}$ and $\g$.
\hfill {\small$\Box$}

\medskip

\noindent \textbf{Proof of Proposition \ref{chasym}}

A direct computation shows that for any $\psi\in L^2$:
\begin{equation}\label{calculcha}
\Cha\ket{\psi}\bra{\psi}\Cha=\ket{\Cha\psi}\bra{\Cha\psi}.
\end{equation}

By Theorem \ref{structure}, for $\mu\in\sigma(\g)\cap(0,1)$, there exist $N\in\mathbb{N}$ and $N$ orthogonal planes $\Pi^1_\mu,\ldots\Pi^N_\mu$ such that
\[
E^{\g^2}_{\mu^2}=E^\g_\mu\overset{\perp}{\oplus}E^\g_{-\mu}=\underset{1\le j\le N}{\bigoplus}\Pi^j_\mu,
\]
where each plane is $\g$-invariant with $\g|_{\Pi_\mu}=\ket{v_-}\bra{v_-}-\ket{e_-}\bra{e_-}$ with $Pv_=v_-$ and $\PP e_-=e_-$. The expression of its eigenvectors $f_+$ and $f_-$ are written in \eqref{eigg}, where $e_+\in\text{Ran}\,\PPP$ is chosen such that $v_-=\cos(\theta)e_-+\sin(\theta)e_+$.

As $\Cha$ is \emph{isometric}, then necessarily $E^{\g^2}_{\mu^2}$ is $\Cha$-invariant, and $\Cha \Pi^j_\mu$ is some plane $\wt{\Pi}^{j}_\mu$ in $E^{\g^2}_{\mu^2}$, $\g$-\emph{invariant} (there holds $\mu=\sin(\widehat{\mathbb{C}v_{-},\mathbb{C}e_{-}})$). Let us show that $ \Pi^j_\mu\neq \wt{\Pi}^{j}_\mu$. Indeed, using \eqref{eigg} this would imply that $\Cha e_-=e^{i\phi_1}e_+$ and $\Cha e_+=e^{i\phi_2}e_-$ for some $\phi_1,\phi_2\in\mathbb{R}$ and
\[
-(\ket{\Cha e_-}\bra{\Cha e_+}+\ket{\Cha e_+}\bra{\Cha e_-})=\ket{e_-}\bra{e_+}+\ket{e_+}\bra{e_-}.
\]
In particular there would hold $-e^{i(\phi_1-\phi_2)}=1$ that is $\phi_1-\phi_2\equiv \pi [2\pi]$. However $\Cha$ is an involution so $\Cha^2 e_+=e_+$ and $e^{i(\phi_1-\phi_2)}e_+=e_+$: this gives $\phi_1-\phi_2\equiv 0 [2\pi]$ and contradicts the previous result. 

Thus the two planes are different and the $4$-dimensional space $V_\mu$ they span is $\Cha$ and $\g$-invariant: $E^{\g^2}_{\mu^2}=V_\mu\overset{\perp}{\oplus} W_\mu$. By induction over $\text{Dim}\,W_\mu$, we get that $2N$ is divisible by $4$, that is $N$ is even. We obtain $\tfrac{N}{2}$ such $V_\mu$, written $V_\mu^{j}$. 

In each $V_\mu^j$, let $u_j^a\perp u_j^b$ be two unitary eigenvectors associated to $\mu$. Thus $\Cha u_j^a\perp \Cha u_j^b$ are two eigenvectors associated to $-\mu$. We use Theorem \ref{structure} to decompose $V_\mu^j=\Pi^a\overset{\perp}{\oplus}\Pi^b$ with
\[
\begin{array}{|l}
\forall\,\star\in\{a,b\},\ \Pi_\star=\mathrm{Span}(u_j^\star,u_{-j}^\star)=\mathrm{Span}(e_j^\star,e_{-j}^\star)\\
\g u_{\pm j}^\star=\pm \mu u_{\pm}^\star,\ \mathcal{P}^0_{\mp}e^\star_{\pm j}=0.
\end{array}
\]
We may assume \eqref{eigg} holds for both planes. Our aim is to prove that up to a phase, $\Cha u_{\pm j}^a=u_{\mp}^b$. \emph{A priori} there exist $\phi_0,\phi_1,\phi_2,\theta\in[-\pi,\pi)$ such that
\[
\begin{array}{|l}
\Cha u_j^a=e^{i\phi_1}\cos(\theta)u_{-j}^a+e^{i\phi_2}\sin(\theta)u_{-j}^b,\\
\Cha u_j^b=-e^{i(\phi_1+\phi_0)}\sin(\theta)u_{-j}^a+e^{i(\phi_2+\phi_0)}\cos(\theta)u_{-j}^b.
\end{array}
\]
We may assume $\cos(\theta),\sin(\theta)>0$. Using \eqref{eigg}, and writing $\ov{\phi}_k=\phi_k+\phi_0$, $k\in\{1,2\}$, we get
\[
\begin{array}{r|l}
\Cha e_j^a=-e^{i\phi_1}\cos(\theta)e_{-j}^a-e^{i\phi_2}\sin(\theta)e_{-j}^b,&\Cha e_j^b=e^{i\ov{\phi}_1}\sin(\theta)e_{-j}^a-e^{i\ov{\phi}_2}\cos(\theta)e_{-j}^b,\\
\Cha e_{-j}^a=e^{i\phi_1}\cos(\theta)e_j^a+e^{i\phi_2}\sin(\theta)e_j^b,&\Cha e_{-j}^b=-e^{i\ov{\phi}_1}\sin(\theta)e_j^a+e^{i\ov{\phi}_2}\cos(\theta)e_j^b.
\end{array}
\]
Applying $\Cha$ to $\Cha e_j^a$ we get
\[
e_j^a=e^{i(\ov{\phi}_1-\phi_2)}\big(\sin(\theta)^2-e^{i(\phi_2-\ov{\phi}_1)}\cos(\theta)^2\big)e_j^a-e^{i\phi_0}\frac{\sin(2\theta)}{2}\big( e^{i(\phi_2-\ov{\phi}_1)}+1\big)e_j^b.
\]
Thus $\sin(\theta)=1$ and $\ov{\phi}_1-\phi_2\equiv 0 [2\pi]$. This gives: 
\begin{equation}
\begin{array}{l}
E^{\g^2}_{\mu^2}=\underset{1\le j\le \tfrac{N}{2}}{\overset{\perp}{\oplus}}V_{\mu}^j\text{\ and\ }V_\mu^j=\Pi_{\mu,j}^a\overset{\perp}{\oplus}\Cha \Pi_{\mu,j}^a,
\end{array}
\end{equation}
where each $\Pi_{\mu,j}^a$ and $\Cha \Pi_{\mu,j}^a$ is a spectral plane described in Theorem \ref{structure}.
\hfill {\small$\Box$}


\bibliographystyle{plain}
{\small\bibliography{bibliothese.bib}}

\end{document}